\documentclass[twocolumn,preprintnumbers,amsmath,amssymb,floatfix,superscriptaddress,nofootinbib]{revtex4}

\usepackage{graphicx}
\usepackage{dcolumn}
\usepackage{bm}
\usepackage{hyperref}
\usepackage{color}
\usepackage{fontawesome} 
\usepackage{comment}
\usepackage{slashed}
\usepackage{tikz}
\usetikzlibrary{positioning,decorations.pathmorphing}
\usepackage{braket}
\usepackage{xspace}

\definecolor{deepblue}{rgb}{0.2,0.2,0.8}
\definecolor{deepred}{rgb}{0.8,0.2,0.2}
\definecolor{deeporange}{rgb}{0.8,0.5,0.2}
\definecolor{deepgreen}{rgb}{0.2,0.8,0.2}

\newcommand{\vect}[1]{\boldsymbol{\mathbf{#1}}}
\newcommand{\half}{\frac{1}{2}}

\newcommand{\dd}{{\rm d}}

\definecolor{linkcolor}{rgb}{0.7752941176470588, 0.22078431372549023, 0.2262745098039215}

\newcommand{\nbicon}{{\color{linkcolor}\faFileCodeO}\xspace}
\newcommand{\nblink}[1]{\href{https://github.com/kenvantilburg/luminous-basin/blob/main/code/#1}{\nbicon}}
\newcommand{\githubmaster}{\href{https://github.com/kenvantilburg/luminous-basin/}{\faGithub}\xspace}

\begin{document}

\newcommand{\be}{\begin{equation}}
\newcommand{\ee}{\end{equation}}

\title{First indirect detection constraints on axions in the Solar basin}

\author{William DeRocco}
\email{wderocco@stanford.edu}
\affiliation{Santa Cruz Institute for Particle Physics, University of California, Santa Cruz, CA 95062, USA}

\author{Shalma Wegsman}
\email{srw9487@nyu.edu}
\affiliation{Center for Cosmology and Particle Physics, Department of Physics, New York University, New York, NY 10003, USA}

\author{Brian Grefenstette}
\email{bwgref@srl.caltech.edu}
\affiliation{Cahill Center for Astronomy and Astrophysics, California Institute of Technology, Pasadena, CA 91125, USA}

\author{Junwu Huang}
\email{jhuang@perimeterinstitute.ca}
\affiliation{Perimeter Institute for Theoretical Physics, Waterloo, Ontario N2L 2Y5, Canada}

\author{Ken Van Tilburg}
\email{kenvt@nyu.edu,kvantilburg@flatironinstitute.org}
\affiliation{Center for Cosmology and Particle Physics, Department of Physics, New York University, New York, NY 10003, USA}
\affiliation{Center for Computational Astrophysics, Flatiron Institute, New York, NY 10010, USA}

\date{\today}

\begin{abstract}
Axions with masses of order keV can be produced in great abundance within the Solar core. The majority of Sun-produced axions escape to infinity, but a small fraction of the flux is produced with speeds below the escape velocity. Over time, this process populates a basin of slow-moving axions trapped on bound orbits. These axions can decay to two photons, yielding an observable signature. We place the first limits on this solar basin of axions using recent quiescent solar observations made by the NuSTAR X-ray telescope. We compare three different methodologies for setting constraints, and obtain world-leading limits for axions with masses between 5 and 30 keV, in some cases improving on stellar cooling bounds by more than an order of magnitude in coupling.
\end{abstract}

\maketitle

\section{Introduction}
\label{sec:intro}

The Sun is an exquisite laboratory for new physics. Studies of particles produced in the Sun have provided us with a wealth of information about light, weakly-interacting particles, including neutrinos~\cite{RevModPhys.60.297,PhysRevD.17.2369,1985YaFiz..42.1441M}, axions~\cite{Raffelt:1996wa,Raffelt:1987im,Raffelt:2006cw}, and dark photons \cite{An:2013yfc,Redondo:2013lna,Hardy:2016kme}. Most of these studies have focused solely on the flux of new particles that escape the gravitational potential of the Sun. However, a tiny fraction will be produced at sufficiently small velocities to become gravitationally trapped, forming a ``Solar basin" of new particles. Ref.~\cite{van2020stellar} showed that despite a small injection rate, the local density in this basin can grow to an appreciable fraction of the local dark matter density, thus providing a new target for direct detection experiments.

If these particles are unstable, however, their density in the Solar basin ceases to grow when decays balance injections, roughly when the Sun's age is comparable to the particle's lifetime. Once this steady-state is reached, the basin decays yield a constant flux of decay products. In the case of keV-mass axion-like particles~\cite{axion1,axion2,axion3,Arvanitaki:2009fg,Svrcek:2006yi} (hereafter ``axions'') 
with a coupling to photons, these decays appear as a narrow line in the X-ray band with a characteristic angular distribution. (Certain aspects of these phenomena were foreseen in Ref.~\cite{dilella2003observational}, which attempted to address the solar coronal heating problem with a Kaluza-Klein spectrum of gravitationally-trapped axions.)

This signal is therefore best probed by X-ray telescopes capable of direct solar observations, such as NuSTAR~\cite{NuSTAR:2013yza}. 
While not initially designed as a solar observatory, it has made observations of both the active Sun~\cite{10.1093/mnras/stab2283,Glesener_2017,Duncan_2021} and, more recently, the quiescent Sun~\cite{Marsh_2017,2020SPD....5121013P} during its nine-year lifetime. Having been constructed with point sources in mind, it boasts a sub-arcminute angular resolution and energy range of $3$--$78\,\mathrm{keV}$. 

\begin{figure}
    \centering
    \includegraphics[width=\columnwidth]{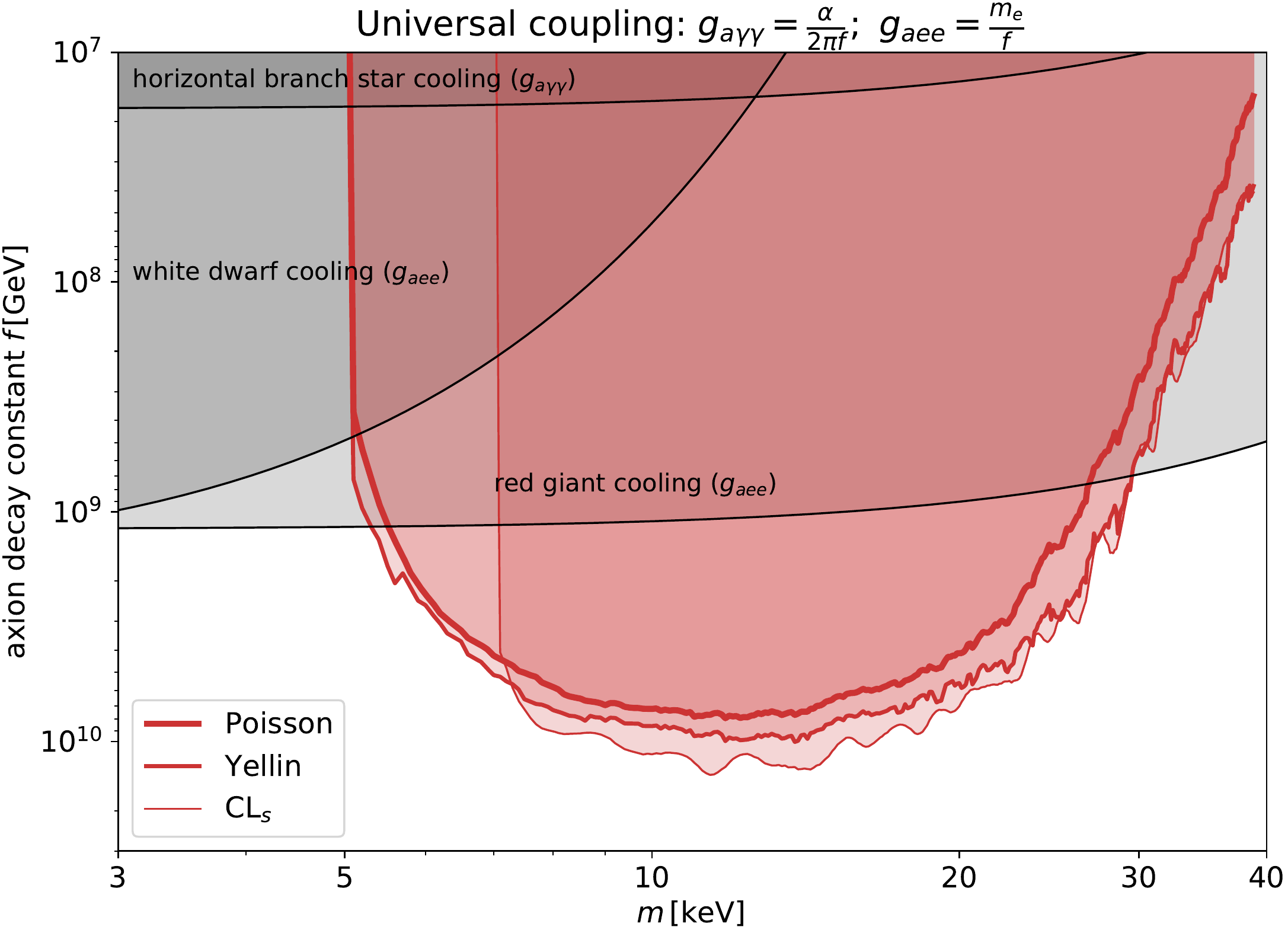}
    \caption{New limits (90{\%}CL) on axion parameter space, with couplings determined by the axion decay constant $f$ as $g_{a\gamma\gamma} = \alpha/2\pi f$ and $g_{aee} = m_e/f$. All regions shaded in red are excluded by our analysis. The thick red curve shows the Poisson limit (Sec.~\ref{subsec:poisson}), the regular red curve shows the optimum cuboid (Yellin) method (Sec.~\ref{subsec:yellin}), and the thin red curve shows the $\mathrm{CL}_s$ limit, which is the most stringent in most of the parameter space (Sec.~\ref{subsec:cls}). This suggests that the modeled smooth X-ray background dominates over the background from microflares in the data used in this analysis. The gray shaded regions are excluded by cooling of horizontal branch stars~\cite{ayala2014revisiting}, white dwarfs~\cite{bertolami2014revisiting}, and red giants~\cite{viaux2013neutrino}. Note that the weakening of the CL$_s$ limit at low masses is due to the use of a 4 keV lower photon energy cutoff to avoid solar contamination in comparison to the 3 keV cutoffs for Poisson and Yellin. \nblink{nb_16_lim_data.ipynb}}
    \label{fig:universal}
\end{figure}

\begin{figure}
\includegraphics[width = 0.5 \textwidth]{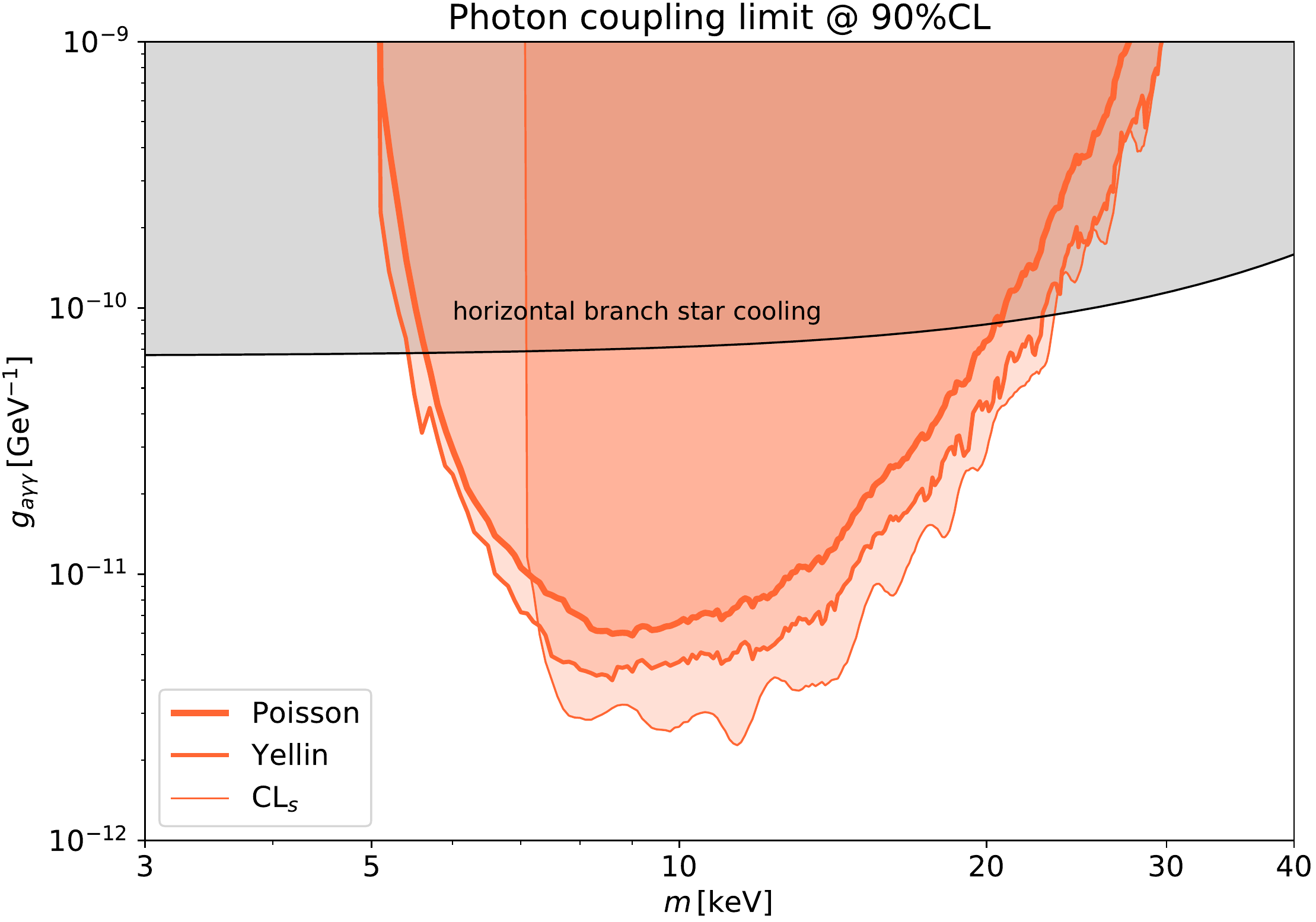}
\caption{Limits on the axion-photon coupling $g_{a\gamma\gamma}$ with $g_{aee} = 0$. See Fig.~\ref{fig:universal} for details. \nblink{nb_16_lim_data.ipynb} } \label{fig:photon}
\end{figure}

In this Letter, we use a subset of NuSTAR's recent quiescent solar limb dwells to constrain the axion-photon and axion-electron coupling, leveraging the characteristic spatial and spectral features of the basin signal. We compute the full production rate for a variety of production processes, and set limits using three different methodologies in order to adopt the most robust bound on parameter space. Two of these methods (Poisson and $\mathrm{CL}_s$ limits) are familiar, but we also present the first multidimensional implementation of a background-agnostic limit-setting methodology based upon the work of Yellin~\cite{yellin2002,yellin2007}. We place constraints stronger than existing limits~\cite{ParticleDataGroup:2018ovx,ayala2014revisiting,Capozzi:2020cbu,Giannotti:2017hny}, by about an order of magnitude, on the axion-photon and axion-electron couplings, for axion masses between $5\,\mathrm{keV}$ and $30\,\mathrm{keV}$. Figures~\ref{fig:universal} and~\ref{fig:photon} show our main results, taken along slices of parameter space given by  $g_{a\gamma\gamma} = \alpha/2\pi f$ and $g_{aee} = m_e/f$ in Fig.~\ref{fig:universal} where $f$ is the axion decay constant, and $g_{a e e} = 0$ in Fig.~\ref{fig:photon}. 

The data and code used to obtain the results of this study is available on GitHub \githubmaster, and a link (\nbicon) below each figure provides the code with which it was generated. We employ natural units wherein $\hbar = c = k_B = 1$.

\section{Signal}
\label{sec:signal}

We consider the leading axion couplings to electrons and photons:
\begin{align}
\mathcal{L} &\subset \half \partial_\mu a \partial^\mu a -\half m^2 a^2\\
&\phantom{\subset}+ \frac{g_{aee}}{2m_e}(\partial_\mu a)\bar{\psi}_e\gamma^\mu \gamma^5 \psi_e - \frac{1}{4} g_{a\gamma\gamma} a F_{\mu\nu}\widetilde{F}^{\mu\nu},
\end{align}
where $m$ is the axion mass, $a$ is the axion field, $g_{aee}$ is the axion coupling to electrons (contributing primarily to axion production), and $g_{a\gamma\gamma}$ is the axion coupling to photons (allowing decays, and also contributing to production).

The energy density injection rate into the basin at a radius $R$ as a result of these processes has been computed in general in Ref.~\cite{van2020stellar}, and can be put into the form 
\begin{align}
	\dot \rho_\mathrm{b}(R) \simeq \frac{3}{16 \pi} \frac{G M_\odot}{R^4} \int \dd^3 R'\,
	\widetilde Q(R') \sqrt{|\Phi(R')|},
	\label{eq:rhodot}
\end{align}
where $M_\odot$ is the mass of the Sun and $\Phi(R')$ is the gravitational potential at the site of production $R'$, with the integral evaluated over the solar volume ($0 < R < R_\odot$). The function $\widetilde{Q}(R')$ encapsulates the production rate of non-relativistic particles into bound orbits~\cite{van2020stellar} and is the sum of three primary components: 
Compton production ($\widetilde{Q}_\mathrm{C}$), bremsstrahlung ($\widetilde{Q}_\mathrm{B}$), and Primakoff production ($\widetilde{Q}_\mathrm{P}$). Expressions for the first two of these components appear in Ref.~\cite{van2020stellar}, and we derive the Primakoff expression 
\begin{equation}
        \widetilde{Q}_\mathrm{P} = g_{a\gamma \gamma}^2 \frac{\sqrt{2} \alpha}{\pi^2} \frac{n_e T}{m_e} \frac{m^4}{e^{m/T}-1}
\end{equation}
in App.~\ref{app:prim}, where $n_e$ is the electron number density, $T$ the temperature, and $\alpha$ the fine structure constant. The quantity $\dot{\rho}_\mathrm{b}(R)$ scales quadratically with $g_{aee}$ (for the $\widetilde{Q}_\mathrm{C}$ and $\widetilde{Q}_\mathrm{B}$ terms) and $g_{a\gamma\gamma}$ (for the $\widetilde{Q}_\mathrm{P}$ term); a plot of its dependence on $m$ can be found in App.~\ref{app:signal}. 

The axions in the basin decay to photons ($a \to 2 \gamma$) either directly via their tree-level photon coupling or at loop level via their electron coupling, at a rate:
\begin{align}
\Gamma_\mathrm{rad} &= \frac{g_{a\gamma \gamma}^2 m^3}{64 \pi}  + \frac{g_{aee}^2 \alpha^2 m^7}{9216 \pi^3 m_e^6} \label{eq:Gammarad}\\
&\approx 2.38 \times 10^{-1} \, \mathrm{Gyr}^{-1} \left( \frac{g_{a\gamma\gamma}}{10^{-12} \, \mathrm{GeV}^{-1}}\right)^2 \left(\frac{m}{10\,\mathrm{keV}}\right)^3 \nonumber \\
&\phantom{\approx} + 5.02 \times 10^{-11} \, \mathrm{Gyr}^{-1} \left( \frac{g_{aee}}{10^{-13}}\right)^2 \left(\frac{m}{10\,\mathrm{keV}}\right)^7 \nonumber.
\end{align}
If the axion's lifetime $\Gamma_\mathrm{rad}^{-1}$ is shorter than the Sun's age $t_\odot = 4.6~\mathrm{Gyr}$, then the system reaches a steady state in which the rate of basin energy density injection ($\dot{\rho}_\mathrm{b}$, which we take to be time-independent) is balanced by the losses due to axion decay ($\rho_\mathrm{b} \Gamma_\mathrm{rad}$). In general, the present-day signal flux is proportional to $\rho_\mathrm{b}\Gamma_\mathrm{rad}|_{t=t_{\odot}} = \dot{\rho}_\mathrm{b}(1-\exp(-\Gamma_{\mathrm{rad}} t_\odot))$. We are primarily interested in the region $R_\odot < R < 2 R_\odot$, where the majority of solar basin axions reside, and where secular perturbations from the planets can be neglected~\cite{giovanetti2022orbital}. In App.~\ref{app:signal}, we argue re-absorption of basin particles can also be ignore for the parameter space considered. 

Since axions trapped in the solar basin have low velocities ($v < v_\mathrm{esc}$), the decay to two photons takes place effectively at rest in the frame of the observer, so each photon acquires an energy $E_\gamma \simeq m/2$. The spectral signature is therefore a line at X-ray energies.

The angular signature follows from the characteristic $1/R^4$ dependence of the basin injection rate, a universal feature of injection into a $1/R$ potential~\cite{van2020stellar}. Integrating along a non-Sun-crossing line of sight yields a $1/\theta^3$ fall-off in observed flux with increasing $\theta$, where $\theta$ is the angle from the center of the Sun. Additionally, there is a doubling discontinuity of the signal at the solar limb ($\theta_\odot \equiv \arcsin[R_\odot/\mathrm{AU}]$), where decays obscured by the Sun for $\theta < \theta_\odot$ become visible when $\theta > \theta_\odot$. A plot of the resulting angular template can be found in App.~\ref{app:signal}. 

Explicitly, we compute the number flux of decay photons per unit solid angle $\dd\Omega$ at Earth by integrating along the line of sight distance $z$:
\begin{align}
\frac{\dd N}{\dd t \, \dd A \, \dd \Omega} &=\int_0^{z_*} \dd z \, \frac{z^2}{4\pi z^2} \frac{\rho_\mathrm{b} \Gamma_\mathrm{rad}}{E_\gamma} \\
&= \frac{\dot{\rho}_\mathrm{b}(R_\odot)}{2\pi m} R_\odot^4 \left[ 1- \exp\lbrace - \Gamma_\mathrm{rad} t_\odot \rbrace \right] \int_0^{z_*} \dd z \frac{1}{R^4} \nonumber,
\end{align}
with $z_* \simeq \mathrm{AU} \big(\cos \theta -\sqrt{\sin^2 \theta_\odot - \sin^2 \theta} \big)$ for $\theta<\theta_\odot$, and $z_* = \infty$ for $\theta>\theta_\odot$. We neglect the $\sim10^{-3}~R_\odot$-thick region above the solar surface in which the optical depth drops abruptly to zero as it is negligible in comparison to the spatial resolution of the detector.
Rewriting the radius in terms of the line-of-sight distance and angle, $R = \mathrm{AU} \sqrt{\sin^2 \theta + (\cos \theta - z/\mathrm{AU})^2}$, we can extract the angular template function $T(\theta) \propto \int \dd z \, R^{-4} $, normalized such that $T(0) = 1$. The signal flux can then be put in the more convenient form
\begin{align}
 &\hspace{-1em}\frac{\dd N}{\dd t \, \dd A \, \dd \Omega} \equiv S_0 T(\theta) =  \frac{\dot{\rho}_\mathrm{b}(R_\odot) R_\odot}{6\pi m} \left[1-e^{-\Gamma_\mathrm{rad} t_\odot}\right] T(\theta) \label{eq:S0}\\
&\approx\frac{2.75 \times 10^{-7}}{\mathrm{s} \, \mathrm{cm}^{2} \, \mathrm{arcsec}^{2}} \left[\frac{\dot{\rho}_\mathrm{b}(R_\odot)/m}{10^{12} \, \mathrm{cm}^{-3}\,\mathrm{Gyr}^{-1}}\right]\left[1-e^{-\Gamma_\mathrm{rad} t_\odot}\right]T(\theta), \nonumber
\end{align}
with $S_0$ the signal flux per unit solid angle at $\theta = 0$. The total number of expected signal events $\mu$ is the expression of Eq.~\ref{eq:S0} integrated over the exposure (corrected by livetime), ancillary response functions (ARFs), and field of view.

\section{Data and Analysis}
\label{sec:analysis}

\subsection{Data processing}
\label{sec:data}

The data used for our analysis were collected by NuSTAR on Sep 12, 2020 during a series of quiescent limb dwells. Our data set is taken from the dwell with the least contamination from localized flares (Orbit 2) and further restricted to that orbit's combination of camera head units (CHUs) with maximal livetime (CHU12). (It is necessary to restrict the data to a particular combination of CHUs in order to ensure the consistency of the spatial coordinates in our analysis; when pointing at the Sun, CHU4 is blinded, hence an uncertainty of $2\,\mathrm{arcmin}$ is introduced on the relative pointing direction between different CHU combinations~\cite{nudata}.) Our selection comprises about $1500\,\mathrm{s}$ of observations, during which the solar center underwent a $1.26\,\mathrm{arcmin}$ shift through NuSTAR's field of view. All events in NuSTAR's calibrated energy range ($3$--$78\,\mathrm{keV}$) were recorded. (A spatial plot of the data can be found in Appendix~\ref{app:realdata}.) 

The photon collection efficiency of NuSTAR over the field of view is quantified in a discrete $13\times13$ partition of subfields for both detectors (A and B), each with angular extent of $\mathrm{arcmin}\times\mathrm{arcmin}$ and separate ancillary response function (ARF) computed using the extended-source functionality of the \texttt{nuproducts} pipeline~\cite{nudata} under the assumption of a spatially uniform background. These ARFs are effective collection areas that take into account detector effects such as aperture stop obscuration, detector absorption, and vignetting~\cite{nudata} that vary over the field of view.

\subsection{Limits}
\label{sec:limits}
To set constraints, we choose three separate methodologies, each with their own associated benefits and challenges. The first is a simple Poisson limit, the second is a generalization of the optimum interval method of Refs.~\cite{yellin2002,yellin2007} (to our knowledge, the first multi-dimensional implementation of the algorithm over binned data), and the third is a likelihood-based $\mathrm{CL}_s$ method.

\subsubsection{Poisson limit}
\label{subsec:poisson}

The premise behind the Poisson limit is simple: the expected signal counts should not exceed the \textit{total} recorded counts at some level of confidence. We identify the signal region by integrating over the observation time and spatial coordinates, but restricting to a narrow window in energy $E\in [m/2-2\sigma_E, m/2+2\sigma_E]$, where $\sigma_E = 0.166$ keV is the spectral resolution. (See App.~\ref{app:energyres}). 
Denoting by $N$ the total number of recorded counts in that energy interval, and by $\mu$ the corresponding expected number of counts in a given signal model specified by $(m,g_{a\gamma \gamma},g_{aee}, \alpha_\odot, \delta_\odot)$, the Poisson likelihood is given by 
\be
    P(N| m,g_{a\gamma \gamma},g_{aee}, \alpha_\odot, \delta_\odot) = \frac{\mu^{N} e^{-\mu}}{N!}.
\ee
The $90\%$ confidence limit $\mu_\text{lim}$ is set at the value of $\mu$, which depends on $(m,g_{a\gamma \gamma},g_{aee}, \alpha_\odot, \delta_\odot)$, such that the cumulative probability for $\mu > \mu_\text{lim}$ is less than $0.1$. In practice, we compute a Poisson limit on $S_0$ (Eq.~\ref{eq:S0}) for each $(m ,\alpha_\odot, \delta_\odot)$, and conservatively report the least stringent limit among a set of 317 discrete, uniformly-spaced solar positions $(\alpha_\odot, \delta_\odot)$ within $6\,\mathrm{arcmin}$ around the fiducial solar position. This simple limit-setting procedure suffers from the loss of spatial and temporal information, and sets the standard for the methods of Secs.~\ref{subsec:yellin} and~\ref{subsec:cls}.

\subsubsection{Yellin limit}
\label{subsec:yellin}

The ``Poisson'' upper limit we obtained in Sec.~\ref{subsec:poisson} stems from a comparison between the expected signal events and the observed events integrated over the entire field of view and observation time. However, the signal has a known spatial, spectral, and temporal dependence, which one can exploit to mitigate some of the background contamination, even for a completely unknown distribution of the background, as proposed by Yellin in Refs.~\cite{yellin2002,yellin2007}.
The method is based on selecting regions within the observed range that contain exceptionally few events in comparison to the signal expectation, or equivalently, exceptionally large regions that contain a given number of events. By construction, this method is relatively insensitive to parts of the signal region with high background (e.g.~a flare). 
Yellin's method~\cite{yellin2002} was initially designed for searches using a signal model with a known one-dimensional distribution, for data that contain few events (e.g.~\cite{Henderson:2008bn}). It is relatively straightforward to find the largest interval which contains a certain number of events, from which one can then find the most improbable of these intervals, the ``optimum interval,'' in the data by comparing to Monte Carlo (MC) simulations generated with \emph{only signal events}. Hence, a background-independent limit can be obtained.

In the case of our analysis, the data exists in a four-dimensional space spanned by $(\alpha,\delta) \equiv (\mathrm{RA},\mathrm{DEC})$ coordinates in the field of view, energy, and time. A main background from the Sun comes from solar (micro)flares, which are localized features in both space and time. Therefore, we generalize the optimum interval method to a multi-dimensional space as outlined in Ref.~\cite{yellin2007}, by identifying the ``optimum cuboid'' in the four spacetime dimensions. This dramatically increases the computational complexity, requiring efficient binning and downsampling strategies discussed in App.~\ref{app:yellin} that preserve the power of the optimum cuboid method, while handling the high dimensionality and statistics of our data set. We perform Monte Carlo simulations of pure signal to compute the cumulative distribution functions (CDFs) for the largest cuboid volumes enclosing $n$ events for $n = 0,1,2,\dots,\,N_\mathrm{data}$. These CDFs can then be combined to place a $90\%~\mathrm{CL}$ limit on the signal flux parametrized by $S_0$ in Eq.~\ref{eq:S0}. As for the Poisson limit, we repeat the procedure for 317 different solar positions, and report the least stringent limit for each value of $m$.

The main advantage of this optimum cuboid method is that it is completely independent of any background model, making it ideal for data sets with unknown backgrounds.

\subsubsection{$\mathrm{CL}_s$ method}
\label{subsec:cls}

A (typically) more stringent upper limit can be set with the $\mathrm{CL}_s$ method~\cite{Read_2002}, which requires a sufficiently flexible model capable of capturing the background while not overfitting.
We make no attempt at modeling the intermittent background from solar (micro-)flares.
The dominant non-solar background in most of the energy range of interest arises from cosmic X-rays that enter the detector at a glancing angle, never having passed through the optical bench. Subleading contributions arise from solar lines and internal components of the telescope. We choose to model these three backgrounds using the spectral shapes measured in Ref.~\cite{Wik:2014boa}, but allow their respective normalizations to float; a plot of the spectra is shown in App.~\ref{app:likelihood}. We treat these backgrounds as spatially uniform, unaffected by the ARFs, since they are not focused by the X-ray optics. Additionally, we raise the minimum energy cutoff for the data to $4\,\mathrm{keV}$ to avoid contamination from solar activity. See Appendix~\ref{app:background} for more detail on the background modeling procedure. 

The likelihood can be expressed as a product of Poisson likelihoods over spectral and spatial bins. We add a Gaussian prior around the fiducial solar position with standard deviation $\sigma_\odot = 2\,\mathrm{arcmin}$. (See Appendix~\ref{app:likelihood} for the full expression.) For each axion mass $m$, the model parameters are therefore the signal flux $S_0$, the initial solar position, and the normalizations of the three background components. 
We perform Markov Chain Monte Carlo (MCMC) analyses at each $m$, and then apply the $\mathrm{CL}_s$ method~\cite{Read_2002} on the resulting marginalized posterior for $S_0$ to place our constraint.

\section{Results}
\label{sec:results}

\begin{figure}
\includegraphics[width = 0.5 \textwidth]{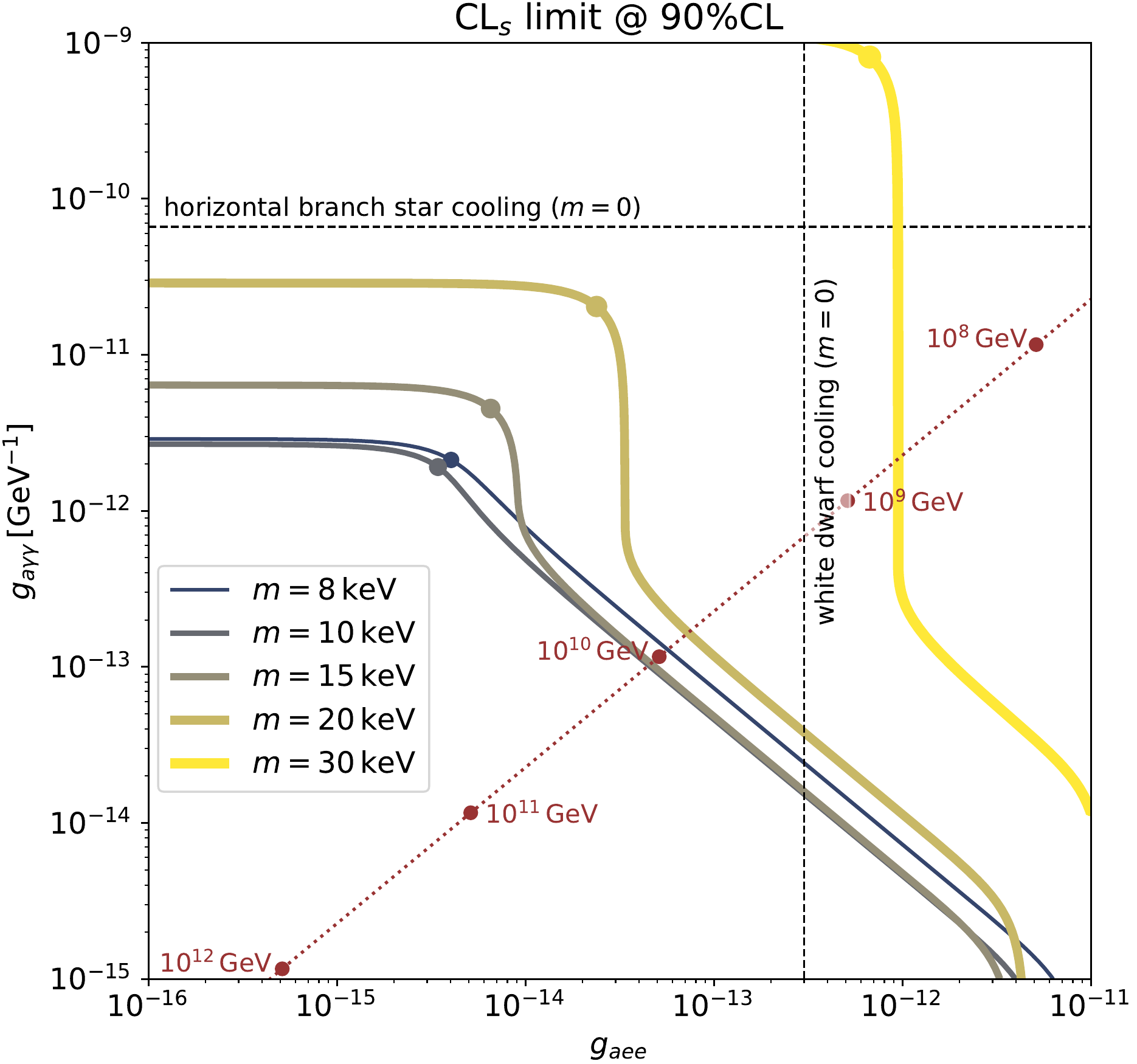}
\caption{The $\mathrm{CL}_s$ limits in the full three-dimensional parameter space, plotted on axes of $g_{a\gamma\gamma}$ and $g_{aee}$, with contours denoting different axion masses $m$. The red dotted line is the slice corresponding to the ``universal coupling'' displayed in Fig.~\ref{fig:universal}. The horizontal and vertical dashed lines correspond to the cooling constraints (at $m=0$) on the axion-photon coupling from horizontal branch stars~\cite{ayala2014revisiting} and on the axion-electron coupling from white dwarfs~\cite{bertolami2014revisiting}, respectively. \nblink{nb_16_lim_data.ipynb}} \label{fig:contours}
\end{figure}

The resulting limits are depicted in Figs.~\ref{fig:universal},~\ref{fig:photon}, and~\ref{fig:contours}. We represent our results in the three-dimensional parameter space spanned by $m, g_{a\gamma\gamma}$, and $g_{aee}$ in a few different ways. The first is to unify the two couplings in terms of an axion decay constant $f$ via the relations $g_{a\gamma\gamma} = \alpha/2\pi f$ and $g_{aee} = m_e/f$, corresponding to an axion with a tree-level electron coupling, and one-loop anomaly coupling to photons, as in DFSZ-type models~\cite{Dine:1981rt,Zhitnitsky:1980tq}. In this case, the axion is produced mainly through the axion-electron coupling $(g_{aee})$, and decays primarily through the axion-photon coupling ($ g_{a\gamma\gamma}$). This yields the limits displayed in Fig.~\ref{fig:universal}, up to one order of magnitude stronger than existing stellar cooling constraints~\cite{ParticleDataGroup:2018ovx,Capozzi:2020cbu,Giannotti:2017hny,ayala2014revisiting,bertolami2014revisiting,viaux2013neutrino}. 

We plot in Fig.~\ref{fig:photon} the limits in $m$--$g_{a\gamma\gamma}$ parameter space, with zero axion-electron coupling, where the axion-photon coupling is responsible for both the production and the decay of the axion. These limits also (approximately) apply to the case where the axion-electron coupling is generated at one-loop order, as in KSVZ-type models~\cite{Kim:1979if,Shifman:1979if}. Our analysis places a limit on an axion coupling exclusively to photons, over an order of magnitude below existing constraints from horizontal branch star cooling~\cite{ayala2014revisiting}.

Finally, we plot our results in $g_{aee}$--$g_{a\gamma\gamma}$ space in Fig.~\ref{fig:contours}, with contours corresponding to different axion masses, encapsulating all information at discrete mass values, and showcasing the scaling of our constraints in various limits. 
Tracing a single curve from the upper left to the lower right, one first sees the region in which the axion production is dominated by the Primakoff process, resulting in a bound independent of the electron coupling, and more stringent than horizontal branch star cooling constraints~\cite{ayala2014revisiting} for all but the highest mass displayed. At the dot, the Compton process overtakes as the dominant production mode, and our constraint is approximately independent of $g_{a\gamma\gamma}$. At even weaker photon couplings, the axion lifetime becomes longer than the age of the Sun, hence the basin decay has not yet fully reached detailed balance with the injection rate (see Eqs.~\ref{eq:Gammarad} and~\ref{eq:S0}), weakening the constraint to a contour at fixed product $g_{aee}^2 g_{a\gamma\gamma}^2$. 
Finally, at very small values of the photon coupling (lower right), the limits become vertical again, indicating the onset of axion decay via an electron loop, occurring well into the region constrained by white dwarf cooling~\cite{bertolami2014revisiting} (vertical dashed line).  
The red dotted line corresponds to the ``universal coupling'' slice from Fig.~\ref{fig:universal}.

The limits from the optimal cuboid (Yellin) and likelihood ($\mathrm{CL}_s$) methods outperform the simple Poisson limit over the entire parameter space, as the former two are able to isolate the signal from the background. The optimum cuboid method offers the highest potential gain in data sets with localized, sporadic features such as flares. (In reality, the realized gain is artificially small, as we already picked a highly quiescent solar period.) Both the Yellin and Poisson limits are completely independent of any background modeling. The likelihood limit typically outperforms the optimum cuboid limit since it includes background fitting of the temporally constant cosmic and instrumental X-ray backgrounds.

\section{Conclusion}

In this Letter, we have presented a first search for X-rays from axion decays in the Solar basin, using observations by the NuSTAR telescope.
We set new constraints on axion parameter space by over an order of magnitude beyond existing bounds in the mass range of $5$--$30\,\mathrm{keV}$.

Our analysis can be augmented by including a background model of the quiescent, micro-flaring Sun, which would strengthen the $\mathrm{CL}_s$ limit, and make possible a potential positive detection using the likelihood method. Longer, dedicated solar observations would also improve the sensitivity of both the likelihood and optimum cuboid methods. The optimum cuboid limit could also tighten with the inclusion of more (and possibly noisier) data, as it automatically selects the least likely region of observation space. Both analyses could also be marginally improved with a more precise knowledge of the fiducial position of the Solar center. A calibration of NuSTAR's ARFs down to lower energies would also extend the sensitivity down to about $2\,\mathrm{keV}$.

The Sun is not the only star that can produce an axion basin: compact remnants such as white dwarfs and neutron stars are attractive targets, as are stars with hot, dense cores, such as (super-)giants and horizontal-branch stars. For targets with extremely low backgrounds, the smaller signal fluxes from these more distant sources could be compensated by stacking multiple exposures. Such strategies could extend the first-of-its-kind indirect detection search for a stellar basin presented here to a larger mass range, even smaller couplings, and other weakly-coupled particles, such as dark photons~\cite{lasenby2021dark} and milli-charged particles~\cite{berlin2021helioscope}. As we have shown in this Letter, indirect detection of stellar basins can probe heretofore unexplored parameter space for weakly-coupled particles.
With the plethora of extensions outlined above, stellar basins will be an exciting target in the hunt for new physics beyond the Standard Model.

\medskip

\acknowledgments{\textit{Acknowledgements---}We thank Iain Hannah, Crystal Kim, and David Smith for discussion regarding NuSTAR data products and instrumental response. We are grateful to Robert Lasenby for sharing preliminary results upon which App.~\ref{app:prim} is based, as well as many clarifying discussions and comments on the draft. We thank Asimina Arvanitaki and Matthew Johnson for comments on the draft, and Andrea Caputo, Joshua Foster, Cristina Mondino, Benjamin Safdi, Edoardo Vitagliano, and Huacheng Yu for useful conversations. WD is supported by the U.S.A. Department of Energy, Grant No.\ DE-SC0010107. Research at Perimeter Institute is supported in part by the Government of Canada through the Department of Innovation, Science and Economic Development Canada and by the Province of Ontario through the Ministry of Colleges and Universities. The Center for Computational Astrophysics at the Flatiron
Institute is supported by the Simons Foundation.} 

\bibliography{LuminousBasin}

\begin{thebibliography}{51}
\expandafter\ifx\csname natexlab\endcsname\relax\def\natexlab#1{#1}\fi
\expandafter\ifx\csname bibnamefont\endcsname\relax
  \def\bibnamefont#1{#1}\fi
\expandafter\ifx\csname bibfnamefont\endcsname\relax
  \def\bibfnamefont#1{#1}\fi
\expandafter\ifx\csname citenamefont\endcsname\relax
  \def\citenamefont#1{#1}\fi
\expandafter\ifx\csname url\endcsname\relax
  \def\url#1{\texttt{#1}}\fi
\expandafter\ifx\csname urlprefix\endcsname\relax\def\urlprefix{URL }\fi
\providecommand{\bibinfo}[2]{#2}
\providecommand{\eprint}[2][]{\url{#2}}

\bibitem[{\citenamefont{Bahcall and Ulrich}(1988)}]{RevModPhys.60.297}
\bibinfo{author}{\bibfnamefont{J.~N.} \bibnamefont{Bahcall}} \bibnamefont{and}
  \bibinfo{author}{\bibfnamefont{R.~K.} \bibnamefont{Ulrich}},
  \bibinfo{journal}{Rev. Mod. Phys.} \textbf{\bibinfo{volume}{60}},
  \bibinfo{pages}{297} (\bibinfo{year}{1988}),
  \urlprefix\url{https://link.aps.org/doi/10.1103/RevModPhys.60.297}.

\bibitem[{\citenamefont{Wolfenstein}(1978)}]{PhysRevD.17.2369}
\bibinfo{author}{\bibfnamefont{L.}~\bibnamefont{Wolfenstein}},
  \bibinfo{journal}{Phys. Rev. D} \textbf{\bibinfo{volume}{17}},
  \bibinfo{pages}{2369} (\bibinfo{year}{1978}),
  \urlprefix\url{https://link.aps.org/doi/10.1103/PhysRevD.17.2369}.

\bibitem[{\citenamefont{{Mikheyev} and {Smirnov}}(1985)}]{1985YaFiz..42.1441M}
\bibinfo{author}{\bibfnamefont{S.~P.} \bibnamefont{{Mikheyev}}}
  \bibnamefont{and} \bibinfo{author}{\bibfnamefont{A.~Y.}
  \bibnamefont{{Smirnov}}}, \bibinfo{journal}{Yadernaya Fizika}
  \textbf{\bibinfo{volume}{42}}, \bibinfo{pages}{1441} (\bibinfo{year}{1985}).

\bibitem[{\citenamefont{Raffelt}(1996)}]{Raffelt:1996wa}
\bibinfo{author}{\bibfnamefont{G.~G.} \bibnamefont{Raffelt}},
  \emph{\bibinfo{title}{{Stars as laboratories for fundamental physics}: {The
  astrophysics of neutrinos, axions, and other weakly interacting particles}}}
  (\bibinfo{year}{1996}), ISBN \bibinfo{isbn}{978-0-226-70272-8}.

\bibitem[{\citenamefont{Raffelt and Stodolsky}(1988)}]{Raffelt:1987im}
\bibinfo{author}{\bibfnamefont{G.}~\bibnamefont{Raffelt}} \bibnamefont{and}
  \bibinfo{author}{\bibfnamefont{L.}~\bibnamefont{Stodolsky}},
  \bibinfo{journal}{Phys. Rev. D} \textbf{\bibinfo{volume}{37}},
  \bibinfo{pages}{1237} (\bibinfo{year}{1988}).

\bibitem[{\citenamefont{Raffelt}(2008)}]{Raffelt:2006cw}
\bibinfo{author}{\bibfnamefont{G.~G.} \bibnamefont{Raffelt}},
  \bibinfo{journal}{Lect. Notes Phys.} \textbf{\bibinfo{volume}{741}},
  \bibinfo{pages}{51} (\bibinfo{year}{2008}), \eprint{hep-ph/0611350}.

\bibitem[{\citenamefont{An et~al.}(2013)\citenamefont{An, Pospelov, and
  Pradler}}]{An:2013yfc}
\bibinfo{author}{\bibfnamefont{H.}~\bibnamefont{An}},
  \bibinfo{author}{\bibfnamefont{M.}~\bibnamefont{Pospelov}}, \bibnamefont{and}
  \bibinfo{author}{\bibfnamefont{J.}~\bibnamefont{Pradler}},
  \bibinfo{journal}{Phys. Lett. B} \textbf{\bibinfo{volume}{725}},
  \bibinfo{pages}{190} (\bibinfo{year}{2013}), \eprint{1302.3884}.

\bibitem[{\citenamefont{Redondo and Raffelt}(2013)}]{Redondo:2013lna}
\bibinfo{author}{\bibfnamefont{J.}~\bibnamefont{Redondo}} \bibnamefont{and}
  \bibinfo{author}{\bibfnamefont{G.}~\bibnamefont{Raffelt}},
  \bibinfo{journal}{JCAP} \textbf{\bibinfo{volume}{08}}, \bibinfo{pages}{034}
  (\bibinfo{year}{2013}), \eprint{1305.2920}.

\bibitem[{\citenamefont{Hardy and Lasenby}(2017)}]{Hardy:2016kme}
\bibinfo{author}{\bibfnamefont{E.}~\bibnamefont{Hardy}} \bibnamefont{and}
  \bibinfo{author}{\bibfnamefont{R.}~\bibnamefont{Lasenby}},
  \bibinfo{journal}{JHEP} \textbf{\bibinfo{volume}{02}}, \bibinfo{pages}{033}
  (\bibinfo{year}{2017}), \eprint{1611.05852}.

\bibitem[{\citenamefont{Van~Tilburg}(2020)}]{van2020stellar}
\bibinfo{author}{\bibfnamefont{K.}~\bibnamefont{Van~Tilburg}},
  \bibinfo{journal}{arXiv preprint arXiv:2006.12431}  (\bibinfo{year}{2020}).

\bibitem[{\citenamefont{Weinberg}(1978)}]{axion1}
\bibinfo{author}{\bibfnamefont{S.}~\bibnamefont{Weinberg}},
  \bibinfo{journal}{Phys.Rev.Lett.} \textbf{\bibinfo{volume}{40}},
  \bibinfo{pages}{223} (\bibinfo{year}{1978}).

\bibitem[{\citenamefont{Wilczek}(1978)}]{axion2}
\bibinfo{author}{\bibfnamefont{F.}~\bibnamefont{Wilczek}},
  \bibinfo{journal}{Phys.Rev.Lett.} \textbf{\bibinfo{volume}{40}},
  \bibinfo{pages}{279} (\bibinfo{year}{1978}).

\bibitem[{\citenamefont{Peccei and Quinn}(1977)}]{axion3}
\bibinfo{author}{\bibfnamefont{R.}~\bibnamefont{Peccei}} \bibnamefont{and}
  \bibinfo{author}{\bibfnamefont{H.~R.} \bibnamefont{Quinn}},
  \bibinfo{journal}{Phys.Rev.Lett.} \textbf{\bibinfo{volume}{38}},
  \bibinfo{pages}{1440} (\bibinfo{year}{1977}).

\bibitem[{\citenamefont{Arvanitaki et~al.}(2010)\citenamefont{Arvanitaki,
  Dimopoulos, Dubovsky, Kaloper, and March-Russell}}]{Arvanitaki:2009fg}
\bibinfo{author}{\bibfnamefont{A.}~\bibnamefont{Arvanitaki}},
  \bibinfo{author}{\bibfnamefont{S.}~\bibnamefont{Dimopoulos}},
  \bibinfo{author}{\bibfnamefont{S.}~\bibnamefont{Dubovsky}},
  \bibinfo{author}{\bibfnamefont{N.}~\bibnamefont{Kaloper}}, \bibnamefont{and}
  \bibinfo{author}{\bibfnamefont{J.}~\bibnamefont{March-Russell}},
  \bibinfo{journal}{Phys. Rev. D} \textbf{\bibinfo{volume}{81}},
  \bibinfo{pages}{123530} (\bibinfo{year}{2010}), \eprint{0905.4720}.

\bibitem[{\citenamefont{Svrcek and Witten}(2006)}]{Svrcek:2006yi}
\bibinfo{author}{\bibfnamefont{P.}~\bibnamefont{Svrcek}} \bibnamefont{and}
  \bibinfo{author}{\bibfnamefont{E.}~\bibnamefont{Witten}},
  \bibinfo{journal}{JHEP} \textbf{\bibinfo{volume}{06}}, \bibinfo{pages}{051}
  (\bibinfo{year}{2006}), \eprint{hep-th/0605206}.

\bibitem[{\citenamefont{DiLella and Zioutas}(2003)}]{dilella2003observational}
\bibinfo{author}{\bibfnamefont{L.}~\bibnamefont{DiLella}} \bibnamefont{and}
  \bibinfo{author}{\bibfnamefont{K.}~\bibnamefont{Zioutas}},
  \bibinfo{journal}{Astroparticle physics} \textbf{\bibinfo{volume}{19}},
  \bibinfo{pages}{145} (\bibinfo{year}{2003}).

\bibitem[{\citenamefont{Harrison et~al.}(2013)}]{NuSTAR:2013yza}
\bibinfo{author}{\bibfnamefont{F.~A.} \bibnamefont{Harrison}}
  \bibnamefont{et~al.} (\bibinfo{collaboration}{NuSTAR}),
  \bibinfo{journal}{Astrophys. J.} \textbf{\bibinfo{volume}{770}},
  \bibinfo{pages}{103} (\bibinfo{year}{2013}), \eprint{1301.7307}.

\bibitem[{\citenamefont{Cooper et~al.}(2021)\citenamefont{Cooper, Hannah,
  Grefenstette, Glesener, Krucker, Hudson, White, Smith, and
  Duncan}}]{10.1093/mnras/stab2283}
\bibinfo{author}{\bibfnamefont{K.}~\bibnamefont{Cooper}},
  \bibinfo{author}{\bibfnamefont{I.~G.} \bibnamefont{Hannah}},
  \bibinfo{author}{\bibfnamefont{B.~W.} \bibnamefont{Grefenstette}},
  \bibinfo{author}{\bibfnamefont{L.}~\bibnamefont{Glesener}},
  \bibinfo{author}{\bibfnamefont{S.}~\bibnamefont{Krucker}},
  \bibinfo{author}{\bibfnamefont{H.~S.} \bibnamefont{Hudson}},
  \bibinfo{author}{\bibfnamefont{S.~M.} \bibnamefont{White}},
  \bibinfo{author}{\bibfnamefont{D.~M.} \bibnamefont{Smith}}, \bibnamefont{and}
  \bibinfo{author}{\bibfnamefont{J.}~\bibnamefont{Duncan}},
  \bibinfo{journal}{Monthly Notices of the Royal Astronomical Society}
  \textbf{\bibinfo{volume}{507}}, \bibinfo{pages}{3936} (\bibinfo{year}{2021}),
  ISSN \bibinfo{issn}{0035-8711},
  \eprint{https://academic.oup.com/mnras/article-pdf/507/3/3936/40346090/stab2283.pdf},
  \urlprefix\url{https://doi.org/10.1093/mnras/stab2283}.

\bibitem[{\citenamefont{Glesener et~al.}(2017)\citenamefont{Glesener, Krucker,
  Hannah, Hudson, Grefenstette, White, Smith, and Marsh}}]{Glesener_2017}
\bibinfo{author}{\bibfnamefont{L.}~\bibnamefont{Glesener}},
  \bibinfo{author}{\bibfnamefont{S.}~\bibnamefont{Krucker}},
  \bibinfo{author}{\bibfnamefont{I.~G.} \bibnamefont{Hannah}},
  \bibinfo{author}{\bibfnamefont{H.}~\bibnamefont{Hudson}},
  \bibinfo{author}{\bibfnamefont{B.~W.} \bibnamefont{Grefenstette}},
  \bibinfo{author}{\bibfnamefont{S.~M.} \bibnamefont{White}},
  \bibinfo{author}{\bibfnamefont{D.~M.} \bibnamefont{Smith}}, \bibnamefont{and}
  \bibinfo{author}{\bibfnamefont{A.~J.} \bibnamefont{Marsh}},
  \bibinfo{journal}{The Astrophysical Journal} \textbf{\bibinfo{volume}{845}},
  \bibinfo{pages}{122} (\bibinfo{year}{2017}),
  \urlprefix\url{https://doi.org/10.3847/1538-4357/aa80e9}.

\bibitem[{\citenamefont{Duncan et~al.}(2021)\citenamefont{Duncan, Glesener,
  Grefenstette, Vievering, Hannah, Smith, Krucker, White, and
  Hudson}}]{Duncan_2021}
\bibinfo{author}{\bibfnamefont{J.}~\bibnamefont{Duncan}},
  \bibinfo{author}{\bibfnamefont{L.}~\bibnamefont{Glesener}},
  \bibinfo{author}{\bibfnamefont{B.~W.} \bibnamefont{Grefenstette}},
  \bibinfo{author}{\bibfnamefont{J.}~\bibnamefont{Vievering}},
  \bibinfo{author}{\bibfnamefont{I.~G.} \bibnamefont{Hannah}},
  \bibinfo{author}{\bibfnamefont{D.~M.} \bibnamefont{Smith}},
  \bibinfo{author}{\bibfnamefont{S.}~\bibnamefont{Krucker}},
  \bibinfo{author}{\bibfnamefont{S.~M.} \bibnamefont{White}}, \bibnamefont{and}
  \bibinfo{author}{\bibfnamefont{H.}~\bibnamefont{Hudson}},
  \bibinfo{journal}{The Astrophysical Journal} \textbf{\bibinfo{volume}{908}},
  \bibinfo{pages}{29} (\bibinfo{year}{2021}),
  \urlprefix\url{https://doi.org/10.3847/1538-4357/abca3d}.

\bibitem[{\citenamefont{Marsh et~al.}(2017)\citenamefont{Marsh, Smith,
  Glesener, Hannah, Grefenstette, Caspi, Krucker, Hudson, Madsen, White
  et~al.}}]{Marsh_2017}
\bibinfo{author}{\bibfnamefont{A.~J.} \bibnamefont{Marsh}},
  \bibinfo{author}{\bibfnamefont{D.~M.} \bibnamefont{Smith}},
  \bibinfo{author}{\bibfnamefont{L.}~\bibnamefont{Glesener}},
  \bibinfo{author}{\bibfnamefont{I.~G.} \bibnamefont{Hannah}},
  \bibinfo{author}{\bibfnamefont{B.~W.} \bibnamefont{Grefenstette}},
  \bibinfo{author}{\bibfnamefont{A.}~\bibnamefont{Caspi}},
  \bibinfo{author}{\bibfnamefont{S.}~\bibnamefont{Krucker}},
  \bibinfo{author}{\bibfnamefont{H.~S.} \bibnamefont{Hudson}},
  \bibinfo{author}{\bibfnamefont{K.~K.} \bibnamefont{Madsen}},
  \bibinfo{author}{\bibfnamefont{S.~M.} \bibnamefont{White}},
  \bibnamefont{et~al.}, \bibinfo{journal}{The Astrophysical Journal}
  \textbf{\bibinfo{volume}{849}}, \bibinfo{pages}{131} (\bibinfo{year}{2017}),
  \urlprefix\url{https://doi.org/10.3847/1538-4357/aa9122}.

\bibitem[{\citenamefont{{Paterson} et~al.}(2020)\citenamefont{{Paterson},
  {Hannah}, {Grefenstette}, {Hudson}, {Krucker}, and
  {Glesener}}}]{2020SPD....5121013P}
\bibinfo{author}{\bibfnamefont{S.}~\bibnamefont{{Paterson}}},
  \bibinfo{author}{\bibfnamefont{I.}~\bibnamefont{{Hannah}}},
  \bibinfo{author}{\bibfnamefont{B.}~\bibnamefont{{Grefenstette}}},
  \bibinfo{author}{\bibfnamefont{H.}~\bibnamefont{{Hudson}}},
  \bibinfo{author}{\bibfnamefont{S.}~\bibnamefont{{Krucker}}},
  \bibnamefont{and}
  \bibinfo{author}{\bibfnamefont{L.}~\bibnamefont{{Glesener}}}, in
  \emph{\bibinfo{booktitle}{AAS/Solar Physics Division Meeting}}
  (\bibinfo{year}{2020}), vol.~\bibinfo{volume}{52} of
  \emph{\bibinfo{series}{AAS/Solar Physics Division Meeting}}, p.
  \bibinfo{pages}{210.13}.

\bibitem[{\citenamefont{Yellin}(2002)}]{yellin2002}
\bibinfo{author}{\bibfnamefont{S.}~\bibnamefont{Yellin}},
  \bibinfo{journal}{Phys. Rev. D} \textbf{\bibinfo{volume}{66}},
  \bibinfo{pages}{032005} (\bibinfo{year}{2002}),
  \urlprefix\url{https://link.aps.org/doi/10.1103/PhysRevD.66.032005}.

\bibitem[{\citenamefont{Yellin}(2007)}]{yellin2007}
\bibinfo{author}{\bibfnamefont{S.}~\bibnamefont{Yellin}},
  \emph{\bibinfo{title}{Extending the optimum interval method}}
  (\bibinfo{year}{2007}), \eprint{0709.2701}.

\bibitem[{\citenamefont{Tanabashi et~al.}(2018)}]{ParticleDataGroup:2018ovx}
\bibinfo{author}{\bibfnamefont{M.}~\bibnamefont{Tanabashi}}
  \bibnamefont{et~al.} (\bibinfo{collaboration}{Particle Data Group}),
  \bibinfo{journal}{Phys. Rev. D} \textbf{\bibinfo{volume}{98}},
  \bibinfo{pages}{030001} (\bibinfo{year}{2018}).

\bibitem[{\citenamefont{Ayala et~al.}(2014)\citenamefont{Ayala, Dom{\'\i}nguez,
  Giannotti, Mirizzi, and Straniero}}]{ayala2014revisiting}
\bibinfo{author}{\bibfnamefont{A.}~\bibnamefont{Ayala}},
  \bibinfo{author}{\bibfnamefont{I.}~\bibnamefont{Dom{\'\i}nguez}},
  \bibinfo{author}{\bibfnamefont{M.}~\bibnamefont{Giannotti}},
  \bibinfo{author}{\bibfnamefont{A.}~\bibnamefont{Mirizzi}}, \bibnamefont{and}
  \bibinfo{author}{\bibfnamefont{O.}~\bibnamefont{Straniero}},
  \bibinfo{journal}{Physical review letters} \textbf{\bibinfo{volume}{113}},
  \bibinfo{pages}{191302} (\bibinfo{year}{2014}).

\bibitem[{\citenamefont{Capozzi and Raffelt}(2020)}]{Capozzi:2020cbu}
\bibinfo{author}{\bibfnamefont{F.}~\bibnamefont{Capozzi}} \bibnamefont{and}
  \bibinfo{author}{\bibfnamefont{G.}~\bibnamefont{Raffelt}},
  \bibinfo{journal}{Phys. Rev. D} \textbf{\bibinfo{volume}{102}},
  \bibinfo{pages}{083007} (\bibinfo{year}{2020}), \eprint{2007.03694}.

\bibitem[{\citenamefont{Giannotti et~al.}(2017)\citenamefont{Giannotti,
  Irastorza, Redondo, Ringwald, and Saikawa}}]{Giannotti:2017hny}
\bibinfo{author}{\bibfnamefont{M.}~\bibnamefont{Giannotti}},
  \bibinfo{author}{\bibfnamefont{I.~G.} \bibnamefont{Irastorza}},
  \bibinfo{author}{\bibfnamefont{J.}~\bibnamefont{Redondo}},
  \bibinfo{author}{\bibfnamefont{A.}~\bibnamefont{Ringwald}}, \bibnamefont{and}
  \bibinfo{author}{\bibfnamefont{K.}~\bibnamefont{Saikawa}},
  \bibinfo{journal}{JCAP} \textbf{\bibinfo{volume}{10}}, \bibinfo{pages}{010}
  (\bibinfo{year}{2017}), \eprint{1708.02111}.

\bibitem[{\citenamefont{Bertolami et~al.}(2014)\citenamefont{Bertolami,
  Melendez, Althaus, and Isern}}]{bertolami2014revisiting}
\bibinfo{author}{\bibfnamefont{M.~M.} \bibnamefont{Bertolami}},
  \bibinfo{author}{\bibfnamefont{B.~E.} \bibnamefont{Melendez}},
  \bibinfo{author}{\bibfnamefont{L.~G.} \bibnamefont{Althaus}},
  \bibnamefont{and} \bibinfo{author}{\bibfnamefont{J.}~\bibnamefont{Isern}},
  \bibinfo{journal}{Journal of Cosmology and Astroparticle Physics}
  \textbf{\bibinfo{volume}{2014}}, \bibinfo{pages}{069} (\bibinfo{year}{2014}).

\bibitem[{\citenamefont{Viaux et~al.}(2013)\citenamefont{Viaux, Catelan,
  Stetson, Raffelt, Redondo, Valcarce, and Weiss}}]{viaux2013neutrino}
\bibinfo{author}{\bibfnamefont{N.}~\bibnamefont{Viaux}},
  \bibinfo{author}{\bibfnamefont{M.}~\bibnamefont{Catelan}},
  \bibinfo{author}{\bibfnamefont{P.~B.} \bibnamefont{Stetson}},
  \bibinfo{author}{\bibfnamefont{G.}~\bibnamefont{Raffelt}},
  \bibinfo{author}{\bibfnamefont{J.}~\bibnamefont{Redondo}},
  \bibinfo{author}{\bibfnamefont{A.~A.} \bibnamefont{Valcarce}},
  \bibnamefont{and} \bibinfo{author}{\bibfnamefont{A.}~\bibnamefont{Weiss}},
  \bibinfo{journal}{Physical Review Letters} \textbf{\bibinfo{volume}{111}},
  \bibinfo{pages}{231301} (\bibinfo{year}{2013}).

\bibitem[{\citenamefont{Giovanetti et~al.}(2022)\citenamefont{Giovanetti,
  Lasenby, and Van~Tilburg}}]{giovanetti2022orbital}
\bibinfo{author}{\bibfnamefont{C.}~\bibnamefont{Giovanetti}},
  \bibinfo{author}{\bibfnamefont{R.}~\bibnamefont{Lasenby}}, \bibnamefont{and}
  \bibinfo{author}{\bibfnamefont{K.}~\bibnamefont{Van~Tilburg}},
  \bibinfo{journal}{in preparation}  (\bibinfo{year}{2022}).

\bibitem[{\citenamefont{Perri}()}]{nudata}
\bibinfo{author}{\bibfnamefont{M.~e.~a.} \bibnamefont{Perri}},
  \emph{\bibinfo{title}{The nustar data analysis software guide}},
  \bibinfo{note}{\url{http://heasarc.gsfc.nasa.gov/docs/nustar/analysis/nustar_swguide.pdf}}.

\bibitem[{\citenamefont{Henderson et~al.}(2008)\citenamefont{Henderson, Monroe,
  and Fisher}}]{Henderson:2008bn}
\bibinfo{author}{\bibfnamefont{S.}~\bibnamefont{Henderson}},
  \bibinfo{author}{\bibfnamefont{J.}~\bibnamefont{Monroe}}, \bibnamefont{and}
  \bibinfo{author}{\bibfnamefont{P.}~\bibnamefont{Fisher}},
  \bibinfo{journal}{Phys. Rev. D} \textbf{\bibinfo{volume}{78}},
  \bibinfo{pages}{015020} (\bibinfo{year}{2008}), \eprint{0801.1624}.

\bibitem[{\citenamefont{Read}(2002)}]{Read_2002}
\bibinfo{author}{\bibfnamefont{A.~L.} \bibnamefont{Read}},
  \bibinfo{journal}{Journal of Physics G: Nuclear and Particle Physics}
  \textbf{\bibinfo{volume}{28}}, \bibinfo{pages}{2693} (\bibinfo{year}{2002}),
  \urlprefix\url{https://doi.org/10.1088/0954-3899/28/10/313}.

\bibitem[{\citenamefont{Wik et~al.}(2014)}]{Wik:2014boa}
\bibinfo{author}{\bibfnamefont{D.~R.} \bibnamefont{Wik}} \bibnamefont{et~al.},
  \bibinfo{journal}{Astrophys. J.} \textbf{\bibinfo{volume}{792}},
  \bibinfo{pages}{48} (\bibinfo{year}{2014}), \eprint{1403.2722}.

\bibitem[{\citenamefont{Dine et~al.}(1981)\citenamefont{Dine, Fischler, and
  Srednicki}}]{Dine:1981rt}
\bibinfo{author}{\bibfnamefont{M.}~\bibnamefont{Dine}},
  \bibinfo{author}{\bibfnamefont{W.}~\bibnamefont{Fischler}}, \bibnamefont{and}
  \bibinfo{author}{\bibfnamefont{M.}~\bibnamefont{Srednicki}},
  \bibinfo{journal}{Phys. Lett. B} \textbf{\bibinfo{volume}{104}},
  \bibinfo{pages}{199} (\bibinfo{year}{1981}).

\bibitem[{\citenamefont{Zhitnitsky}(1980)}]{Zhitnitsky:1980tq}
\bibinfo{author}{\bibfnamefont{A.~R.} \bibnamefont{Zhitnitsky}},
  \bibinfo{journal}{Sov. J. Nucl. Phys.} \textbf{\bibinfo{volume}{31}},
  \bibinfo{pages}{260} (\bibinfo{year}{1980}).

\bibitem[{\citenamefont{Kim}(1979)}]{Kim:1979if}
\bibinfo{author}{\bibfnamefont{J.~E.} \bibnamefont{Kim}},
  \bibinfo{journal}{Phys. Rev. Lett.} \textbf{\bibinfo{volume}{43}},
  \bibinfo{pages}{103} (\bibinfo{year}{1979}).

\bibitem[{\citenamefont{Shifman et~al.}(1980)\citenamefont{Shifman, Vainshtein,
  and Zakharov}}]{Shifman:1979if}
\bibinfo{author}{\bibfnamefont{M.~A.} \bibnamefont{Shifman}},
  \bibinfo{author}{\bibfnamefont{A.~I.} \bibnamefont{Vainshtein}},
  \bibnamefont{and} \bibinfo{author}{\bibfnamefont{V.~I.}
  \bibnamefont{Zakharov}}, \bibinfo{journal}{Nucl. Phys. B}
  \textbf{\bibinfo{volume}{166}}, \bibinfo{pages}{493} (\bibinfo{year}{1980}).

\bibitem[{\citenamefont{Lasenby and Van~Tilburg}(2021)}]{lasenby2021dark}
\bibinfo{author}{\bibfnamefont{R.}~\bibnamefont{Lasenby}} \bibnamefont{and}
  \bibinfo{author}{\bibfnamefont{K.}~\bibnamefont{Van~Tilburg}},
  \bibinfo{journal}{Physical Review D} \textbf{\bibinfo{volume}{104}},
  \bibinfo{pages}{023020} (\bibinfo{year}{2021}).

\bibitem[{\citenamefont{Berlin and Schutz}(2021)}]{berlin2021helioscope}
\bibinfo{author}{\bibfnamefont{A.}~\bibnamefont{Berlin}} \bibnamefont{and}
  \bibinfo{author}{\bibfnamefont{K.}~\bibnamefont{Schutz}},
  \bibinfo{journal}{arXiv preprint arXiv:2111.01796}  (\bibinfo{year}{2021}).

\bibitem[{\citenamefont{Raffelt}(1986)}]{Raffelt:1985nk}
\bibinfo{author}{\bibfnamefont{G.~G.} \bibnamefont{Raffelt}},
  \bibinfo{journal}{Phys. Rev. D} \textbf{\bibinfo{volume}{33}},
  \bibinfo{pages}{897} (\bibinfo{year}{1986}).

\bibitem[{\citenamefont{Altherr}(1991)}]{Altherr:1990tf}
\bibinfo{author}{\bibfnamefont{T.}~\bibnamefont{Altherr}},
  \bibinfo{journal}{Annals Phys.} \textbf{\bibinfo{volume}{207}},
  \bibinfo{pages}{374} (\bibinfo{year}{1991}).

\bibitem[{\citenamefont{Altherr and Kraemmer}(1992)}]{Altherr:1992mf}
\bibinfo{author}{\bibfnamefont{T.}~\bibnamefont{Altherr}} \bibnamefont{and}
  \bibinfo{author}{\bibfnamefont{U.}~\bibnamefont{Kraemmer}},
  \bibinfo{journal}{Astropart. Phys.} \textbf{\bibinfo{volume}{1}},
  \bibinfo{pages}{133} (\bibinfo{year}{1992}).

\bibitem[{\citenamefont{Altherr et~al.}(1994)\citenamefont{Altherr,
  Petitgirard, and del Rio~Gaztelurrutia}}]{Altherr:1993zd}
\bibinfo{author}{\bibfnamefont{T.}~\bibnamefont{Altherr}},
  \bibinfo{author}{\bibfnamefont{E.}~\bibnamefont{Petitgirard}},
  \bibnamefont{and} \bibinfo{author}{\bibfnamefont{T.}~\bibnamefont{del
  Rio~Gaztelurrutia}}, \bibinfo{journal}{Astropart. Phys.}
  \textbf{\bibinfo{volume}{2}}, \bibinfo{pages}{175} (\bibinfo{year}{1994}),
  \eprint{hep-ph/9310304}.

\bibitem[{\citenamefont{DeRocco et~al.}(2022)\citenamefont{DeRocco, Galanis,
  and Lasenby}}]{DeRocco:2022rze}
\bibinfo{author}{\bibfnamefont{W.}~\bibnamefont{DeRocco}},
  \bibinfo{author}{\bibfnamefont{M.}~\bibnamefont{Galanis}}, \bibnamefont{and}
  \bibinfo{author}{\bibfnamefont{R.}~\bibnamefont{Lasenby}}
  (\bibinfo{year}{2022}), \eprint{2201.05167}.

\bibitem[{\citenamefont{Phillips et~al.}(2006)\citenamefont{Phillips, Chifor,
  and Dennis}}]{Phillips:2006ne}
\bibinfo{author}{\bibfnamefont{K.~J.~H.} \bibnamefont{Phillips}},
  \bibinfo{author}{\bibfnamefont{C.}~\bibnamefont{Chifor}}, \bibnamefont{and}
  \bibinfo{author}{\bibfnamefont{B.~R.} \bibnamefont{Dennis}},
  \bibinfo{journal}{Astrophys. J.} \textbf{\bibinfo{volume}{647}},
  \bibinfo{pages}{1480} (\bibinfo{year}{2006}), \eprint{astro-ph/0607309}.

\bibitem[{\citenamefont{Grefenstette et~al.}(2016)}]{Grefenstette:2016esh}
\bibinfo{author}{\bibfnamefont{B.~W.} \bibnamefont{Grefenstette}}
  \bibnamefont{et~al.}, \bibinfo{journal}{Astrophys. J.}
  \textbf{\bibinfo{volume}{826}}, \bibinfo{pages}{20} (\bibinfo{year}{2016}),
  \eprint{1605.09738}.

\bibitem[{\citenamefont{{Foreman-Mackey}
  et~al.}(2013)\citenamefont{{Foreman-Mackey}, {Hogg}, {Lang}, and
  {Goodman}}}]{emcee}
\bibinfo{author}{\bibfnamefont{D.}~\bibnamefont{{Foreman-Mackey}}},
  \bibinfo{author}{\bibfnamefont{D.~W.} \bibnamefont{{Hogg}}},
  \bibinfo{author}{\bibfnamefont{D.}~\bibnamefont{{Lang}}}, \bibnamefont{and}
  \bibinfo{author}{\bibfnamefont{J.}~\bibnamefont{{Goodman}}},
  \bibinfo{journal}{PASP} \textbf{\bibinfo{volume}{125}}, \bibinfo{pages}{306}
  (\bibinfo{year}{2013}), \eprint{1202.3665}.

\bibitem[{\citenamefont{Ku et~al.}(1997)\citenamefont{Ku, Liu, and
  Hsu}}]{Ku1997}
\bibinfo{author}{\bibfnamefont{L.-p.} \bibnamefont{Ku}},
  \bibinfo{author}{\bibfnamefont{B.}~\bibnamefont{Liu}}, \bibnamefont{and}
  \bibinfo{author}{\bibfnamefont{W.}~\bibnamefont{Hsu}} (\bibinfo{year}{1997}).

\bibitem[{\citenamefont{Nandy and Bhattacharya}(1998)}]{NANDY199811}
\bibinfo{author}{\bibfnamefont{S.}~\bibnamefont{Nandy}} \bibnamefont{and}
  \bibinfo{author}{\bibfnamefont{B.}~\bibnamefont{Bhattacharya}},
  \bibinfo{journal}{Computers \& Mathematics with Applications}
  \textbf{\bibinfo{volume}{36}}, \bibinfo{pages}{11} (\bibinfo{year}{1998}),
  ISSN \bibinfo{issn}{0898-1221},
  \urlprefix\url{https://www.sciencedirect.com/science/article/pii/S0898122198001254}.

\end{thebibliography}

\appendix

\section{Primakoff production rate}
\label{app:prim}

In this section, we compute the production rate of non-relativistic (NR) axions from the Sun through the axion-photon and axion-electron coupling. Axions can be produced via the Compton and bremsstrahlung processes, as well as the Primakoff and electro-Primakoff processes. The dominant production rate via the axion-electron coupling in the axion mass range of interest is the Compton process (see Ref.~\cite{van2020stellar} for more details), with production function (to be used in Eq.~\ref{eq:rhodot}):
\begin{equation}
        \widetilde{Q}_\mathrm{C} = \frac{\alpha g_{a e e}^2}{2^{3/2}\pi^2} \frac{n_e }{m_e^4} \frac{m^4\sqrt{m^2-\omega_\mathrm{pl}^2}}{e^{m/T}-1},
\end{equation}
where $\omega_\mathrm{pl} \equiv \sqrt{4\pi \alpha n_e / m_e}$ is the plasma frequency.

\begin{figure} [t!]
\begin{center}
\begin{tikzpicture}[line width=1.5 pt, scale=2.5,  photon/.style={decorate,decoration={snake,post length=1mm}}]
 \draw[photon]  (-1.2, 1.0)--(0, 1.0);
 \draw[dashed]  ( 0, 1.0)--(1.2, 1.0);
 \draw[photon] (0, 1.0) --(0, 0.0);
 \draw[solid]    (-1.2,-0.0)--(0, 0.0);
 \draw[solid]    ( 0, 0.0)--(1.2,-0.0);
 \node at (-1.3, 1.1) {$\gamma$};
 \node at ( 1.3, 1.1) {$a$};
 \node at (-1.3,-0.1) {$e$};
 \node at ( 1.3,-0.1) {$e$};
 \node at (-0.6, 1.2) {$K_\gamma = (\omega_{\gamma},\vect{k}_\gamma)$};
 \node at ( 0.6, 1.2) {$\vect{k}_a \approx 0$};
 \node at (-0.6,-0.2) {$\vect{k}_i$};
 \node at ( 0.6,-0.2) {$\vect{k}_f$};
 \node at (0.6, 0.5) {$Q=(q_0,\vect{q})$};
\end{tikzpicture}
\caption{Feynman diagram for the Primakoff process.}
\label{fig:feynman}
\end{center}
\end{figure}
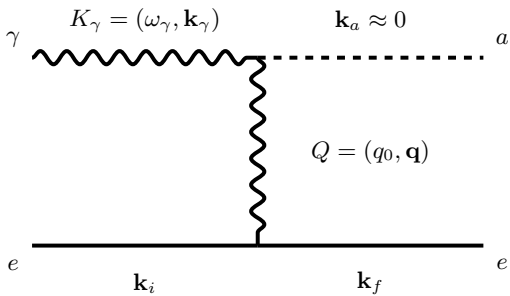

For the axion-photon coupling, the leading production mode is the Primakoff process depicted in Fig.~\ref{fig:feynman}, whose production rate vanishes at leading order in the NR expansion~\cite{Raffelt:1985nk}. At this order, the only 3-momentum in the diagram is the momentum of the photon $\vect{k}_{\gamma} $ and $ \vect{k}_a$, so the matrix element of an $a F \tilde{F}$ coupling, depending on the cross product of $\vect{k}_{\gamma} $ and $ \vect{k}_a$, vanishes as $k_a \to 0$. At subleading order, there are two expansion parameters, the electron velocity $v_e \sim \sqrt{3T/m_e} \sim 0.1$ and the axion velocity $v_a \sim v_{\rm esc} \sim 5 \times 10^{-3}$. In the following, we will compute the Primakoff production rate at $\mathcal{O}(v_e^2)$ in the NR expansion. 

We start by expanding the spinors to first order in the velocity expansion. In the kinematic regime of interest, the incoming photon energy is approximately equal to the axion mass. Only the transverse mode of the photon contributes to the matrix element in vacuum, which is found to be
\begin{equation}
\label{eq:matrixelement}
    \left| \mathcal{M}\right|^2 = \frac{8 g_{a \gamma \gamma}^2 e^2 m^2 (\vect{k}_i\cdot \vect{k}_{\gamma})(\vect{k}_f \cdot \vect{k}_{\gamma})}{\vect{k}_{\gamma}^4 }
\end{equation}
for $m_a \gg \max\lbrace T,\, \omega_\mathrm{pl}\rbrace$, where $\vect{k}_i$ and $\vect{k}_f$ are the initial and final 3-momentum of the electron, and $\vect{k}_{\gamma}$ is the 3-momentum of the incoming photon. 

Extending the result to lower axion masses requires accounting for thermal effects~\cite{Altherr:1990tf,Altherr:1992mf,Altherr:1993zd,DeRocco:2022rze}. 
The production rate of an axion in the limit of vanishing $v_a$ is 
\begin{align}
    \Gamma = &\frac{g_{a \gamma \gamma}^2}{16\pi^2 m}\int^{\infty}_{\omega_p} \dd \omega_{\gamma}\, {k}_{\gamma} \, \frac{\theta(q_0)}{e^{\omega_{\gamma}/T}-1} \\ 
    &\hspace{8em}\times \int^{1}_{-1} \dd (\cos{\phi}) \, \beta(Q) g(K_{\gamma},Q), \nonumber
\end{align}
where $\beta(Q)$ is the imaginary part of the the propagator, $\theta(q_0)$ is a Heaviside theta function, and $ g({K}_{\gamma}, Q)$ is the coupling function:
\begin{align}
    g_\mathrm{L}({K}_{\gamma}, Q) &= Q^2 k_{\gamma}^2\sin^2{\phi} \\
    g_\mathrm{T}({K}_{\gamma}, Q) &= (\omega_{\gamma}^2 q^2 + q_0^2 k_{\gamma}^2) (1+\cos^2{\phi})-4 \omega_{\gamma} q q_0 k_{\gamma} \cos \phi \nonumber
\end{align}
where $g_\mathrm{L}$ and $g_\mathrm{T}$ refer to the coupling functions of longitudinal and transverse plasmons, respectively. The ``Cut'' plasmon 4-momentum is $Q = {(q_0, \vect{q})}$, and that of the ``Pole'' plasmon $K_{\gamma} = {(\omega_{\gamma}, \vect{k}_{\gamma})}$, while $\phi$ is the angle between the 3-momenta $\vect{q}$ and $\vect{k}_{\gamma}$~\cite{Altherr:1990tf}. Following~\cite{Altherr:1990tf}, we refer to the off-shell plasmon with momentum $Q$ (see Fig.~\ref{fig:feynman}) as a ``Cut'' plasmon,  while the on-shell initial state plasmon with momentum $K_{\gamma}$ as a ``Pole'' plasmon. In the following, we will justify the approximations we made to arrive at Eq. \ref{eq:matrixelement} as well as its qualitative features by examining the properties of the functions $ g({K}_{\gamma}, Q)$ and $\beta(Q)$.

The emission rate depends on the coupling function $ g({K}_{\gamma}, Q)$, which contains the Lorentz contraction of the vertices in Fig. \ref{fig:feynman}. The emission of relativistic axions via the Primakoff effect is dominated by the exchange of a longitudinal plasmon. However, the longitudinal coupling function $g_\mathrm{L}$ vanishes in the limit $\phi\rightarrow 0$ ($k_a \rightarrow 0$), signaling a $v_a^2$ suppression as expected (and consistent with the in-vacuum result). In contrast, the transverse coupling function $g_\mathrm{T}$ does not vanish in the same $\phi\rightarrow 0$ limit. In the following, we will therefore focus on the transverse-plasmon exchange. 

The emission rate also depends on $\beta(Q)$, the imaginary part of the propagator of the ``Cut'' plasmon
\begin{equation}
    \beta(Q) = -\frac{2 {\rm Im} \,\pi(Q) \theta(-Q^2)}{[Q^2 -{\rm Re}\,\pi(Q)]^2 +[{\rm Im}\, \pi(Q)]^2},
\end{equation}
where $\pi(Q)$ is part of the polarization tensor $\pi^{\mu\nu}(Q)$ as defined in Ref.~\cite{Altherr:1990tf} and computed in Ref.~\cite{DeRocco:2022rze}:
\begin{align}
    \pi_\mathrm{L}(Q) &= \omega_\mathrm{pl}^2\frac{m_e}{q_0}\xi\left[Z(\xi_+)-Z(\xi_-)\right], \\
    \pi_\mathrm{T}(Q) &=  \omega_\mathrm{pl}^2\left\lbrace1- \left[\sigma^2 - \frac{Q^2}{4 m_e^2}\right]\frac{m_e}{q_0}\xi\left[Z(\xi_+)-Z(\xi_-)\right] \right\rbrace\nonumber,
\end{align}
with $\mathrm{L}$ and $\mathrm{T}$ subscripts referring to longitudinal and transverse plasmons. These results hold for a dilute NR plasma; see Ref.~\cite{DeRocco:2022rze} for an in-depth analysis. In the above, we have employed the definitions $\xi_{\pm} \equiv \xi (1\pm \frac{Q^2}{2 q_0 m_e})$, the plasma dispersion function
\begin{equation}
Z(\xi) = \frac{1}{\sqrt{\pi}} \int^{\infty}_{-\infty} \mathrm{d} x \frac{e^{-x^2}}{x-\xi} = i\sqrt{\pi} e^{-\xi^2}\mathrm{erfc}(- i \xi),    
\end{equation}
with $\xi \equiv q_0 / (q\sigma)$, and $\sigma$ is the electron velocity dispersion~\cite{DeRocco:2022rze}. The plasma dispersion function $Z(\xi)$ contains the entire imaginary part of $\pi(Q)$. We thus find:
\begin{equation}
    \frac{{\rm Im}\,\pi_T(Q)}{{\rm Im}\,\pi_L(Q)} \sim \sigma^2
\end{equation}
for $Q^2 \ll (m_e \sigma)^2$, implying that the emission of relativistic axions through exchange of a transverse plasmon is suppressed by $v_e^2$ compared to the emission through exchange of a longitudinal plasmon, confirming our prior result.

This $\sigma^2$ difference between the longitudinal and transverse plasmon also shows up in the real part of $\pi(Q)$. Whereas the function ${\rm }\,\pi_\mathrm{L}(Q) \rightarrow \omega_\mathrm{pl}^2/\sigma^2 \sim k_D^2$ in the limit $q_0 \rightarrow 0$, the function ${\rm }\,\pi_\mathrm{T}(Q) \rightarrow 0$ in the same limit. Here, $k_D = \sqrt{e^2n_e/T}$ is the Debye screening wave vector. This suggests that the exchange of a transverse plasmon does not experience the same Debye screening as the exchange of a longitudinal plasmon, also as expected.

We can now integrate over the distribution of the initial state electron and photon, as well as the final state phase space to find the production rate of non-relativistic axions as 
\begin{equation}
        \widetilde{Q}_\mathrm{P} = g_{a\gamma \gamma}^2 \frac{\sqrt{2} \alpha}{\pi^2} \frac{n_e T}{m_e} \frac{m^4}{e^{m/T}-1},
\end{equation}
valid for axion masses $0.3 \,{\rm keV} \lesssim \omega_\mathrm{pl} \ll m_a < 2 m_e \sigma \approx 50 \,{\rm keV}$. As $m_a$ approaches $\omega_\mathrm{pl}$, the production rate of non-relativistic axion through the ``Cut-Pole'' diagram becomes exponentially small due to the exponential suppression in the plasma dispersion function at large $\xi$. 

The ``Pole-Pole'' contribution to the NR axion flux can only arise from the decay of the transverse plasmon to a longitudinal plasmon at the same momentum, but this rate is suppressed by the axion velocity $v_a^2$ and will not dominate in the mass range of interest to this work. NR axions can also be produced through the electro-Primakoff effect (``Cut-Cut''). Similarly, the matrix element can also only get contributions at $\mathcal{O}(v_e^2)$, so its rate is always subdominant compared to the Primakoff rate in the Sun.

\section{Basin Evolution and Saturation}
\label{app:signal}

The energy density of axions in the solar basin depends not just on their injection rate from the Compton, bremsstrahlung and Primakoff production processes, but also on their effective accumulation time. There are three  processes that contribute to basin depletion, and therefore set the effective lifetime of an axion in the stellar basin. 

The first process is axion decay, with rate computed in Eq.~\ref{eq:Gammarad}, yielding a radiative lifetime within a few orders of magnitude from the age of the Sun ($t_\odot$) for the parameter space of interest. In the main text (Eq.~\ref{eq:S0} in particular), we have taken this radiative decay as the primary depletion process.

The second process is gravitational ejection, from close encounters or orbital resonances. For our data set based on observations of the solar limb, the vast majority of X-ray photons originate from axions on orbits with aphelia between $R_{\odot}$ and $2R_{\odot}$. For such tight orbits, gravitational ejection times are much longer than the Sun's age~\cite{giovanetti2022orbital}.

\begin{figure}
\includegraphics[width = 0.45 \textwidth]{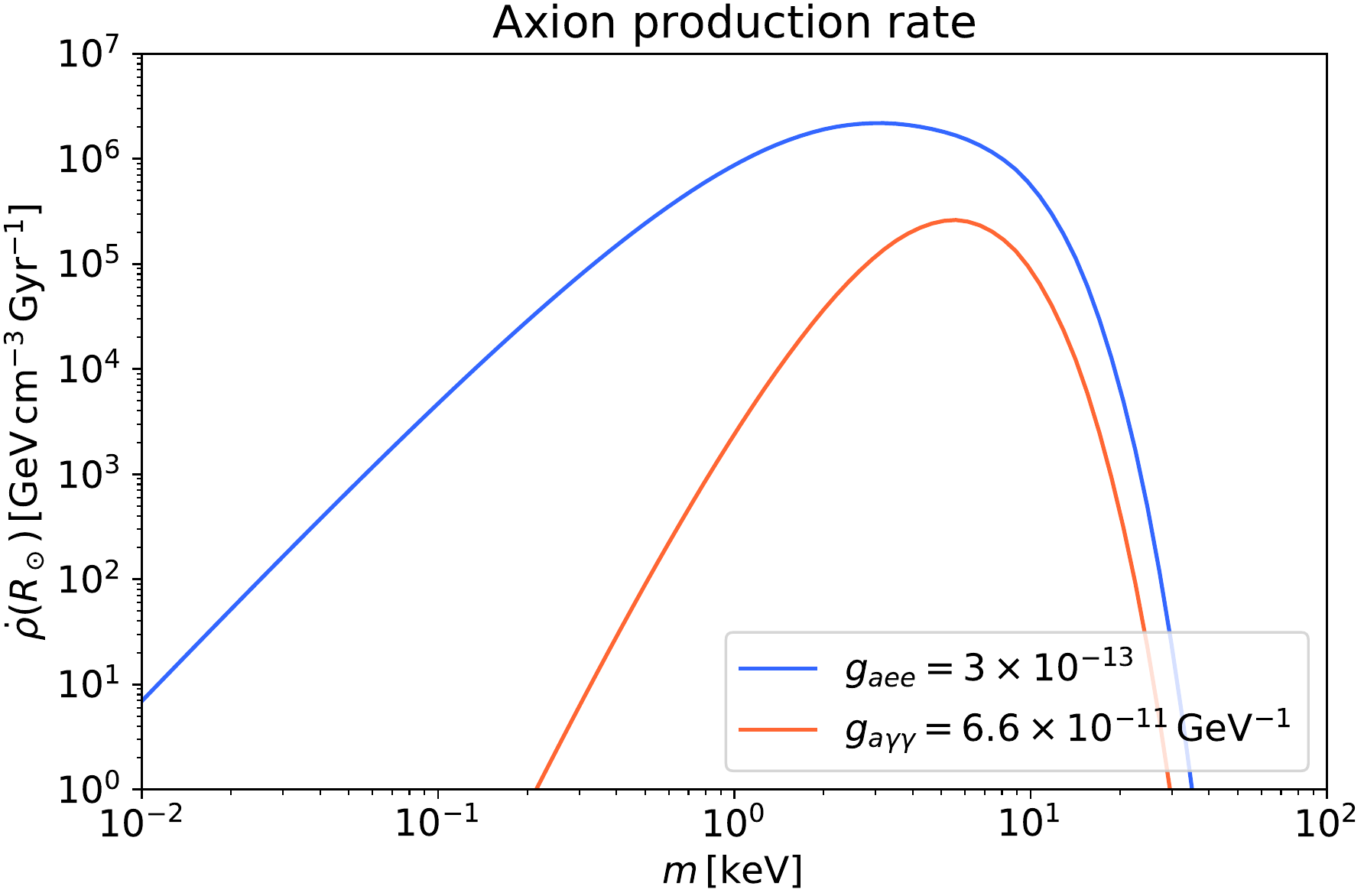}
\includegraphics[width = 0.45 \textwidth]{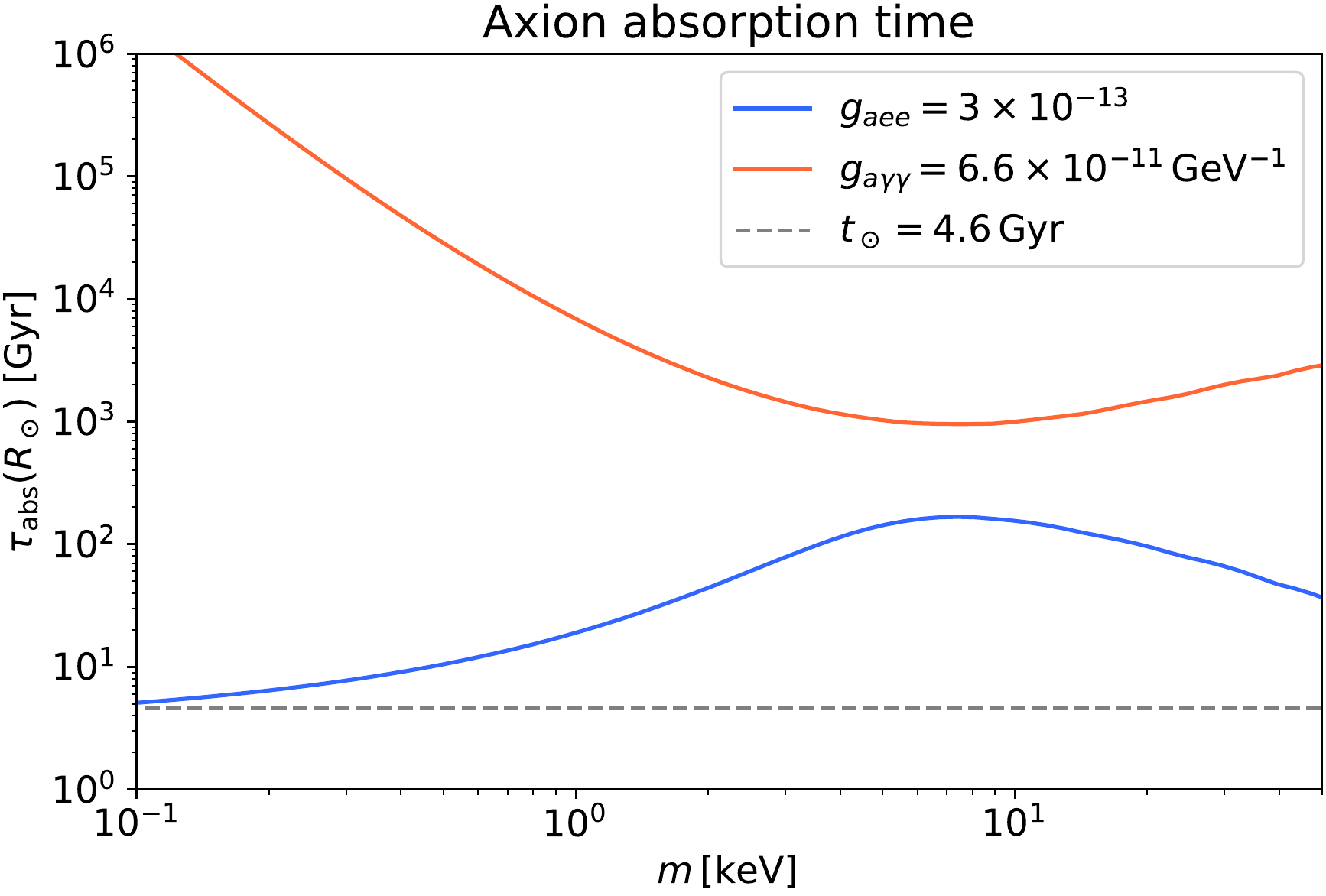}
\caption{\textit{Top panel:} The axion density injection rate at the Solar surface, shown for two representative choices of axion, one coupled only to electrons (blue) and one to only photons (orange), as a function of axion mass. \textit{Bottom panel:} Effective absorption time $\tau_\mathrm{abs}(R)$ at the solar surface ($R = R_\odot$). Even when saturating the stellar cooling bounds for both the $g_{a\gamma\gamma}$ and $g_{aee}$ couplings, the effective absorption time scale is longer than the age of the Sun $t_\odot$, especially for the mass range of interest for the analysis in this work $m \in [5,30]\,\mathrm{keV}]$. \nblink{nb_16_lim_data.ipynb}} \label{fig:rhodot}
\end{figure}

The third process is re-absorption of basin particles in the Sun, which can be estimated as~\cite{van2020stellar}:
\begin{equation}
	\tau_{\rm abs} \simeq \frac{1}{\dot{\rho}_b(R)}m^4 f_{\rm sat} \frac{v_{\rm esc}^3(R)}{6 \pi^2}
	\label{eq:tauabs}
\end{equation}
where $f_{\rm sat} = \langle \frac{1}{e^{m/ T }-1} \rangle$ whose average is weighted by the production function $\widetilde{Q}$ over the solar volume, as in \cite[Eq.~21]{lasenby2021dark}. This time is also much longer than Sun's age for parameter space that is not already excluded for the mass range of interest, as shown in the bottom panel of Fig.~\ref{fig:rhodot}, which assumes $f_\mathrm{sat} = \frac{1}{e^{m/ T_\odot}-1}$ with constant $T_\odot = 1.3 \, \mathrm{keV}$ for simplicity, and evaluates $\tau_{\rm abs}$ at $R = R_\odot$. 

\begin{figure}
\includegraphics[width = 0.45 \textwidth]{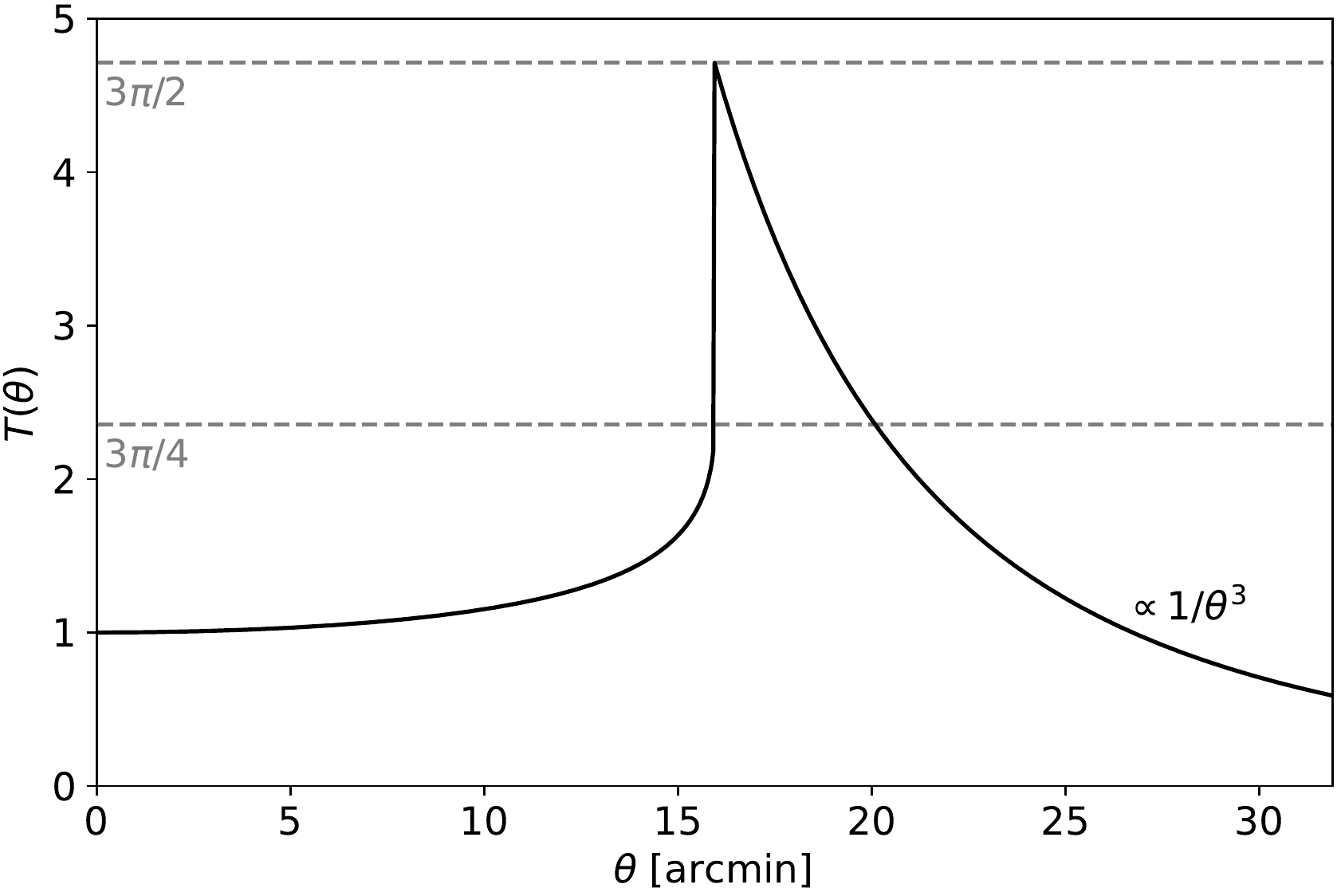}
\caption{Angular dependence of axion decay signal flux, with $\theta$ measured with respect to the center of the Sun. Note the doubling at the solar limb and characteristic $1/\theta^3$ dependence for larger angles, which follows from the $1/R^4$ density profile of the Solar basin.} \label{fig:template}
\end{figure}

\section{Data}
\label{app:data}

\subsection{Observation and detector response}
\label{app:realdata}

We use data recorded by NuSTAR on 12-September-2020 during a series of quiescent limb dwells. Our dataset is taken from the dwell with the least contamination from localized flares (Orbit 2) and further restricted to that orbit's combination of camera head units (CHUs) with maximal livetime (CHU12). We choose to restrict the data to a particular combination of CHUs in order to ensure the consistency of the spatial coordinates in our analysis; Solar observations are performed in NuSTAR's Mode 06 (Science Mode) in which CHU4 is blinded, hence an uncertainty of $\sim$2 arcmin is introduced on the absolute pointing direction between different CHU combinations~\cite{nudata}. This uncertainty also requires our analyses to marginalize over the true solar position. The effective exposure for CHU12 was 1501.17 seconds for focal plane module (FPM) FPMA and 1481.86 seconds for FPMB, during which the solar limb underwent a $1.26'$ shift of ($\Delta$RA, $\Delta$Dec) $ = (0.020^\circ, -0.0084^\circ)$ through NuSTAR's field of view. See Table~\ref{tab:data} for further details of the observation.

\begin{table*}[tp]
    \centering
    \begin{tabular}{|c|c|c|c|c|}
    \hline
         \textbf{Obs ID} & \textbf{Start (UTC)} & \textbf{Stop (UTC)} & \textbf{CALDB version} & \textbf{Exposure (s)}  \\
         \hline
         80610202001 & 2020-09-12T10:21:09 & 2020-09-12T11:56:09 &
         20200912 &
         \begin{tabular}{@{}c@{}}1501.16599845754 (FPMA) \\ 1481.86081041239 (FPMB)\end{tabular}\\
         \hline
    \end{tabular}
    \caption{Metadata of the NuSTAR observations used in this paper.}
    \label{tab:data}
\end{table*}

\begin{figure}[h]
    \centering
    \includegraphics[width = 0.45\textwidth]{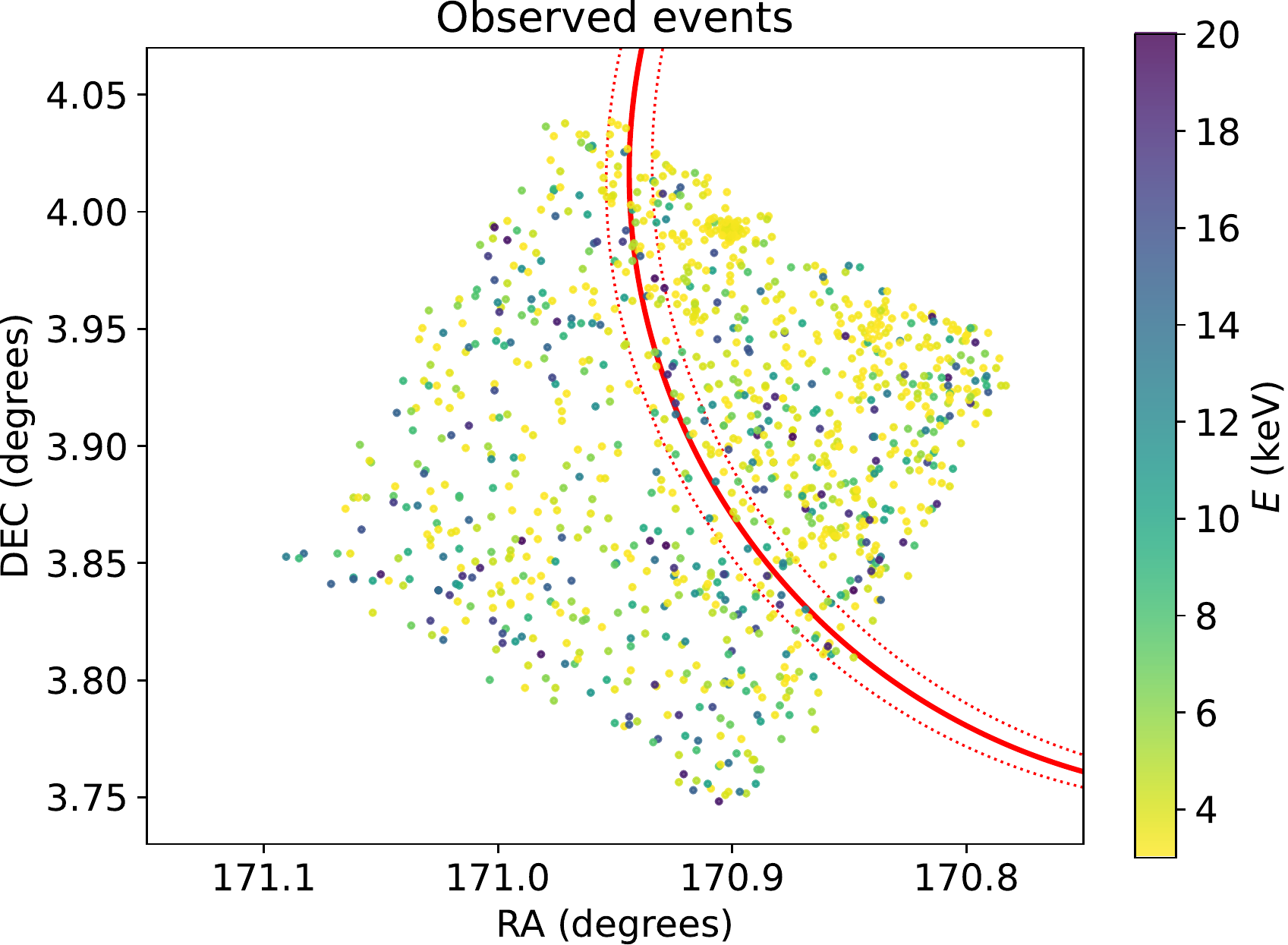}
    \caption{Spatial map of the raw data used in our analysis, restricted to the relevant interval $3$--$20\,\mathrm{keV}$. Each point represents a photon, and is colored according to its energy. The red solid line indicates the edge of the solar disk for the Sun at its fiducial position in the middle of the observation; the dotted red lines depict it at the start and end of the observations, with the Sun moving towards the bottom left. \nblink{nb_12_yellin_proj_data.ipynb}}
    \label{fig:image_raw}
\end{figure}
\begin{figure}[h]
    \centering
    \includegraphics[width = 0.45\textwidth]{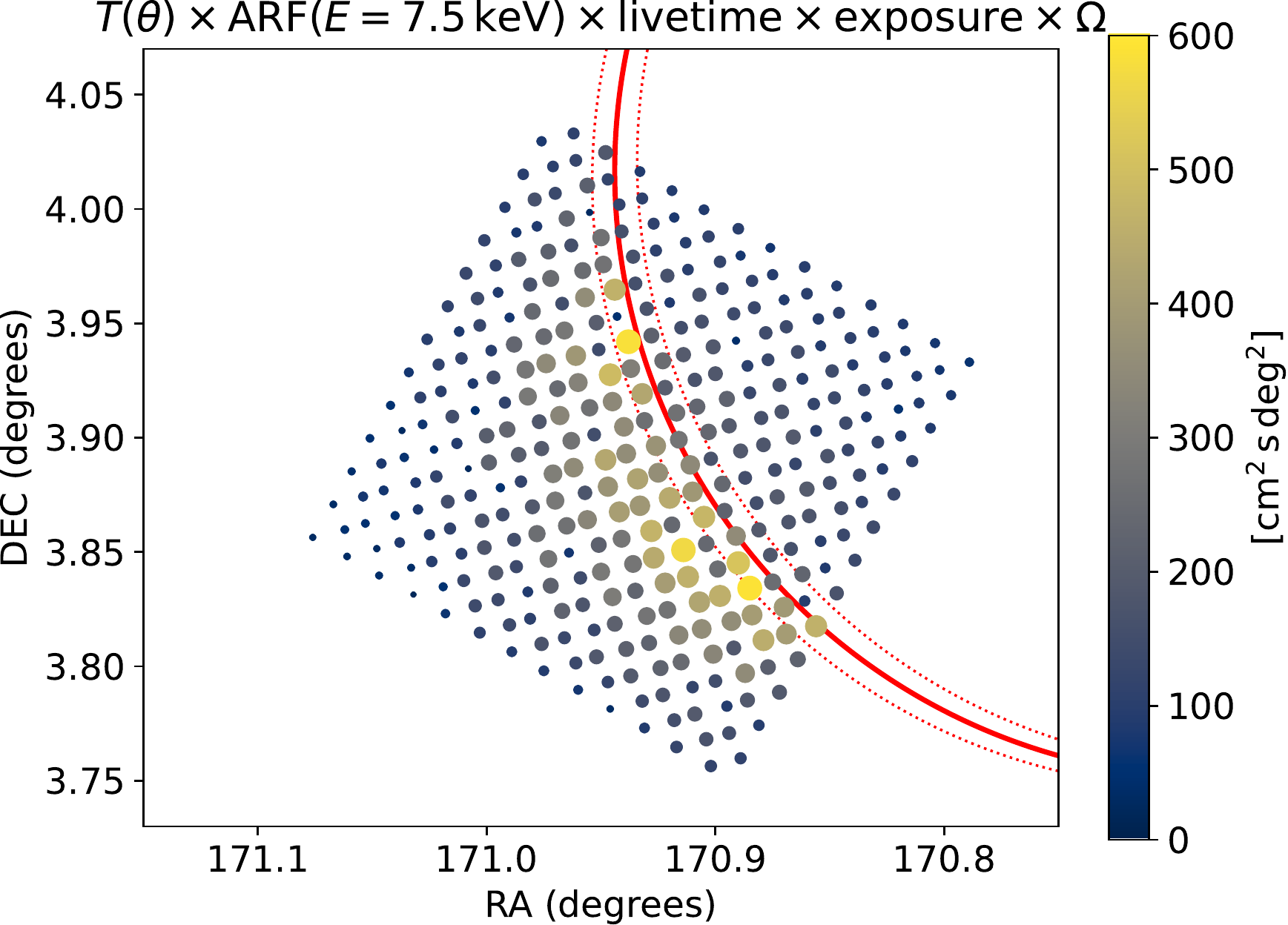}
    \caption{Spatial map proportional to the expected number of counts in the $13\times13$ subfields of both detectors (for $m = 15\,\mathrm{keV}$). Specifically, we plot the product of: $T(\theta)$ with the Sun at its fiducial position in the middle of the observation window; the spatially varying ARFs at $E = 7.5\,\mathrm{keV}$; the effective exposure; the $\Omega = 1\,\mathrm{arcmin}^2$ solid angle of each subfield. \nblink{nb_12_yellin_proj_data.ipynb}}
    \label{fig:image_exp}
\end{figure}

Event data was taken from Level 2 cleaned and calibrated event files. The spatial distribution is shown in Fig.~\ref{fig:image_raw}. Energy was computed from the pulse invariant (PI) using the formula $E = (0.04\,\text{keV})\text{PI} + 1.6\,\text{keV}$. ARFs were computed using the \texttt{nuproducts} pipeline~\cite{nudata}. NuSTAR is primarily a point-source telescope, so the ARFs were computed using the extended source functionality and flat count distribution (assuming spatially uniform background) with no background extraction. Source regions were taken to be the cells associated with the gridding of the FOV into arcmin$^2$ regions ($13 \times 13$ for both detectors) (see Sec.~\ref{sec:data}), with a different ARF calculated separately within each to account for the spatial dependence of the effective area due to detector effects (aperture stop, detector absorption, vignetting). The resulting expected spatial distribution of signal events is shown in Fig.~\ref{fig:image_exp}.

\subsection{Mock data generation}
\label{app:mockdata}

To compare to the true observation data, our limit-setting machinery was tested using mock data for which the axion mass and couplings are known. This mock data was generated in the following manner. 

The background component was generated by sampling uniformly in time over the good time intervals for the observation and sampling energies from an interpolated distribution function formed from the 200-bin histogram of the true data. The background was sampled to be spatially uniform. The number of background counts was sampled according to Poisson statistics with expectation value given by the true number of counts in the data.

The signal component was generated by computing the expected number of counts of the signal model (Eq.~\ref{eq:S0}) within each subgrid of the FOV for which an ARF was computed (taking the Sun to track its fiducial trajectory) and sampling from a Poisson distribution with that mean, then distributing those counts uniformly spatially over the subgrid and spectrally as a Gaussian with width $\sigma_E$ centered at $E = m/2$. The temporal dependence was taken to be uniform. While some approximations made in the generation of this mock data result in slight differences from a true expected signal, the mock data sufficed to verify the efficacy of our modeling framework.

\section{Energy Resolution}
\label{app:energyres}

\begin{figure}
\includegraphics[width = 0.45 \textwidth]{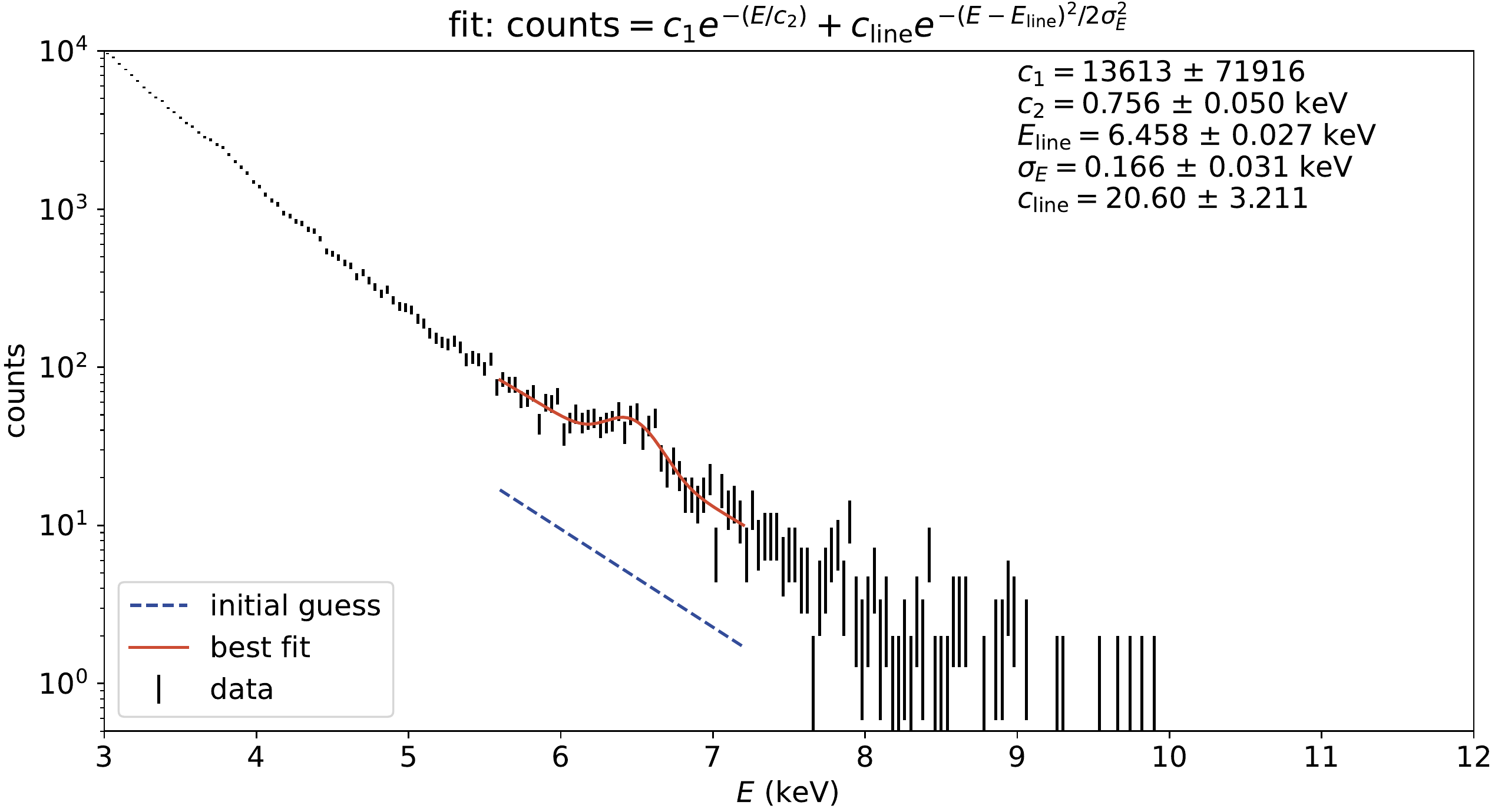}
\caption{Best fit curve to the $6.7\,$keV iron line in observations of the active Sun. The resulting best-fit width of the Gaussian component ($\sigma_E = 0.166$ keV) was taken as the fiducial NuSTAR energy resolution for the entirety of our analysis. \nblink{nb_energy_res.ipynb}} \label{fig:energyres}
\end{figure}

In order to determine a fiducial energy resolution for spectral lines in NuSTAR, we fit a known X-ray line in data of the active Sun with a Gaussian, taking the best fit width of the Gaussian $\sigma_{\text{line}}$ as our fiducial resolution $\sigma_E$.
We target the iron line at $E = 6.7\,\mathrm{keV}$~\cite{Phillips:2006ne} in observations of the active Sun taken on 01-November-2014~\cite[Tab.~1]{Grefenstette:2016esh}), and fit the spectrum with a function of the form
\be
f(E) = c_1 \exp\left(-\frac{E}{c_2}\right) + c_\text{line} \exp\left(-\frac{(E-E_\text{line})^2}{2 \sigma_{\text{line}}^2}\right)
\ee
where $\{c_1, c_2, c_\text{line}, \sigma_{\text{line}}\}$ were simultaneously fit to the data.
The first term is included to model the solar background as a falling exponential, though since the line is narrow, the fit is  insensitive to the particular functional form of this component. The second term models the line itself as a Gaussian centered at an energy $E_{\text{line}}$ with width $\sigma_{\text{line}}$. The results of the fit are shown in Fig.~\ref{fig:energyres}. The fit yields a line energy $E_\text{line} = 6.45\,\mathrm{keV}$ and a width $\sigma_\text{line} = 0.166 \, \mathrm{keV}$, which we choose to use as our fiducial energy resolution $\sigma_E$ for our analyses. This is a known marginal shift in the line energy (which is not yet understood), however the energy resolution is typical for NuSTAR observations \citep[e.g.,][]{NuSTAR:2013yza}.

\section{Likelihood Analysis Method}
\label{app:bglikely}

\subsection{Background characterization}
\label{app:background}

The background present in our data arises from several different sources, as mentioned in the main body of the text. The dominant background arises from cosmic X-rays that enter the detector at a glancing angle, never having passed through the optical bench, however subleading contributions arise from solar lines and internal components of the telescope. We choose to model the background using the spectral shapes measured in~\cite{Wik:2014boa} but allowing the respective normalizations to vary. A plot of the various spectra is shown in Fig.~\ref{fig:spectrum}. $B_1$ (``aperture'' in Fig. 9 of~\cite{Wik:2014boa}) corresponds to the cosmic X-ray background that does not pass through the optical bench, $B_2$ (``int. lines'') to internal lines arising from the radiation environment of the detector, and $B_3$ (``int. cont.'') to gamma-ray Compton scattering and other featureless instrumental components. (See Appendix A of Ref.~\cite{Wik:2014boa} for further discussion of these backgrounds.)

\begin{figure*}
    \centering
    \includegraphics[width = 0.85\textwidth]{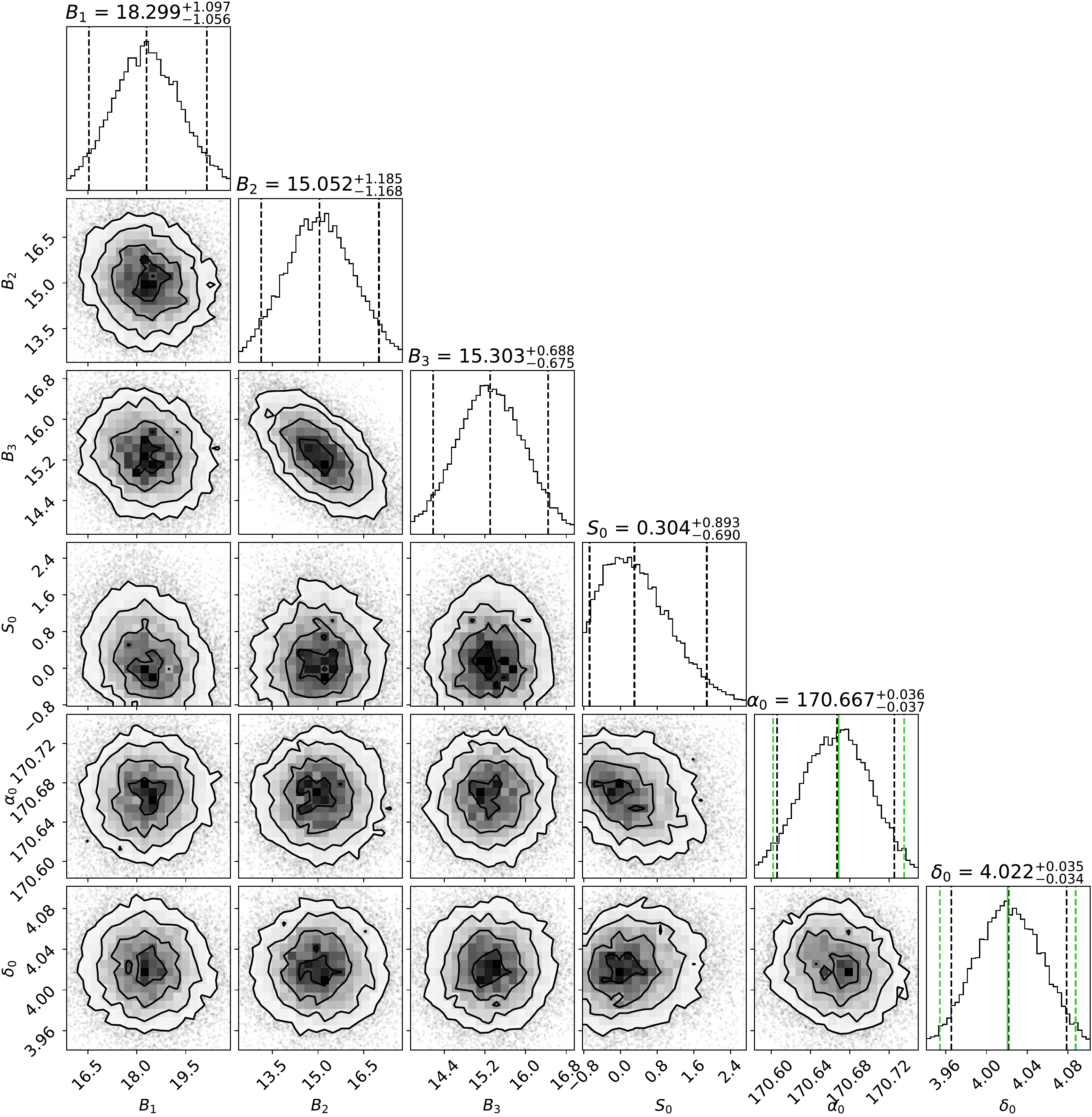}
    \caption{Corner plot for MCMC fitting routine used in CL$_s$ method at $m = 15$ keV. The units are arbitrary for the $B_i$, while $S_0$ is measured in units of $10^{-4}\,\mathrm{cm}^{-2}\,\mathrm{s}^{-1}~\mathrm{deg}^{-2}$, and $\alpha_0$ and $\delta_0$ in degrees. Black (green) dashed lines in the 1D histograms correspond to $5\%$, $50\%$, and $95\%$ quantiles for the posterior (prior). \nblink{nb_15_likelihood_data.ipynb}}
    \label{fig:corner}
\end{figure*}

\begin{figure*}
    \centering
    \includegraphics[width = 0.85\textwidth]{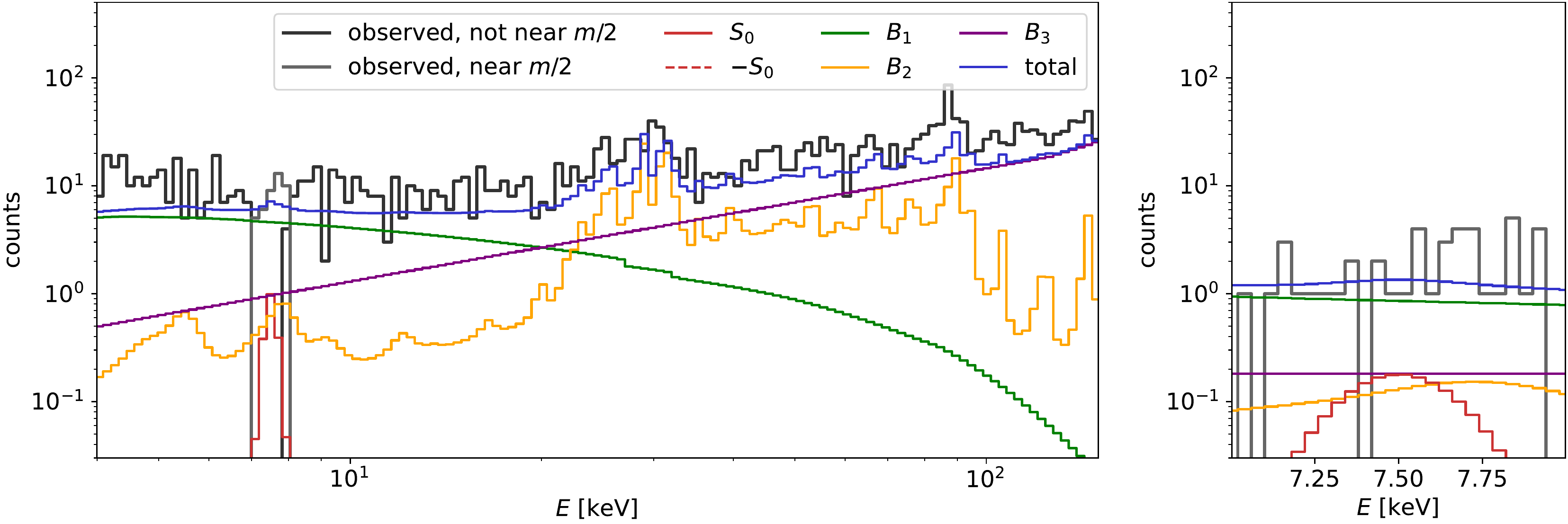}
    \caption{Best fit spectrum as determined during MCMC routine of CL$_s$ limit at $m = 15$ keV. The associated background components are shown separately, with their sum shown in blue. $B_1$ corresponds to the aperture cosmic X-ray background, $B_2$ to internal lines, $B_3$ to internal continuum emission, and $S_0$ to the signal (see App.~\ref{app:background}). The true data are shown in black. \nblink{nb_15_likelihood_data.ipynb}}
    \label{fig:spectrum}
\end{figure*}

There is an additional contribution at very low energies that arises from active regions on the Sun. These stochastic features are not possible to encapsulate in a simple model, hence when performing the CL$_s$ method, we have chosen to excise data contaminated by these microflares by raising the lower energy cutoff on the data used to set this limit. In order to select this cutoff, we performed a global fit of the background model to the observed data for $E_{\text{min}}$ ranging from 3 to $20\,\mathrm{keV}$ with a spacing of $0.24\,\mathrm{keV}$. The $\chi^2$ of the fit and coefficients $B_1, B_2$, and $B_3$ stabilize near $3.8\,\mathrm{keV}$, hence we selected $4\,\mathrm{keV}$ as our $E_{\text{min}}$ for the CL$_s$ analysis. This cutoff explains the dramatic weakening of sensitivity of our CL$_s$ bounds below $m \lesssim 8\,\mathrm{keV}$ in Figs.~\ref{fig:universal} and~\ref{fig:photon}. As can be seen in Fig.~\ref{fig:fit_S0}, with this choice of cutoff, the CL$_s$ method is biased towards finding higher $p$-values, hence is a conservative methodology.

\subsection{Likelihood computation}
\label{app:likelihood}

As stated in Sec.~\ref{app:realdata}, we grid the field of view into $13\times13$ grid of $\mathrm{arcmin}^2$ cells, for both detectors A and B. The data, $\textbf{d}_{ijEt}$, is therefore binned by spatial cell ($i,j = 1,\dots,13$), energy, and time, with energy bin widths of $40\,\mathrm{eV}$. Each of the time bins is multiplied by the active exposure time of that bin. The data is also multiplied by the computed ancillary response function (ARF) that characterizes the detector effective area in each bin and the energy resolution $\sigma_E =0.166 \, \mathrm{keV}$.
The Poisson likelihood for our flux model is a product over all bins:
\begin{align}
    &P(\textbf{d}_{ijEt} | m,g_{a\gamma \gamma},g_{aee}, \alpha_\odot, \delta_\odot, B_1, B_2, B_3) \\
    &\hspace{10em}= \prod_{ijEt} \frac{\mu_{ijEt}^{n_{ijEt}} e^{-\mu_{ijEt}}}{n_{ijEt}!} \nonumber
\end{align}
where $\mu_{ijEt}$ is the expected number of counts in each bin given the flux model, and $n_{ijEt}$ is the observed counts.

The total flux model, including both the expected signal shape and backgrounds (denoted by $f_k$ for functional form and $B_k$ for free normalization, see Sec.~\ref{app:background}), is given by
\begin{align}
    \frac{\dd N}{\dd E \, \dd t} &= 
    B_1 f_{\textrm{aCXB}}(E) 
    + B_2 f_{\textrm{internal}}(E) 
    + B_3 f_{\textrm{cont}}(E) \nonumber \\
    &\phantom{=} + \bigg[ S_0 T(\theta) \frac{e^{-{(E-m/2)^2}/{2\sigma_E^2}}}{\sqrt{2 \pi \sigma_E^2}} \bigg] \textrm{ARF} \times \Omega
\end{align}
where $\Omega = 1\,\mathrm{arcmin}^2$ is the solid angle of each subfield. This leaves six free parameters: $B_1, B_2,$ and $B_3$ representing the background normalizations, $S_0$ for the overall signal strength, and $\alpha_\odot$ and $\delta_\odot$ for the (initial) solar position, which enters $T(\theta)$ through $\theta  = \big \| \big(\alpha,\delta\big) - \big(\alpha_\odot(t),\delta_\odot(t)\big) \big \|$. The angular flux template $T(\theta)$ is derived as shown in Sec.~\ref{sec:signal} and given explicitly by:
\begin{widetext}
\begin{align}
    T(\theta) = \begin{cases} \frac{3\pi}{2} \sin^3 \theta_\odot \csc^3 \theta & \sin \theta_\odot  < \sin \theta \\ 
\frac{3}{4} \sin^3 \theta_\odot \csc^2\theta  
\left\lbrace \frac{4 z_\mathrm{min}}{-1 - 2 z_\mathrm{min}^2 + \cos 2 \theta } + [\pi - 2 \mathrm{arctan}(z_\mathrm{min} \csc \theta) \csc \theta ]\right\rbrace & \sin \theta \le \sin \theta_\odot \end{cases},
\end{align}
\end{widetext}
where $z_\mathrm{min} \equiv \sqrt{\sin^2 \theta_\odot - \sin^2 \theta}$.
The expected signal is thus adjusted for the detector's effective area per bin (via the ARF), and for the motion of the Sun during the data collection.

To set a limit we run a Markov Chain Monte Carlo (MCMC) which maximizes the likelihood function:
\begin{equation}
    \sum_{ijEt} \text{LL}_m + \sum_{ijEt} \text{LL}_{\overline m} + \text{LP}_\odot 
\end{equation}
where $\text{LL}_m$ is the logarithm of the likelihood given the data within $2\sigma_E$ of a given axion mass, and $\text{LL}_{\overline m}$ is that for the data outside of the $2\sigma_E$ range (separated such that only the \emph{energy} spectrum of the non-axion region must be fitted to reduce computation time). We add a Gaussian prior to take into account our imprecise knowledge on the solar position around its fiducial value $(\alpha_\odot^\textrm{fid},\delta_\odot^\textrm{fid})$:
\begin{equation}
     \text{LP}_\odot = -\frac{\cos(\delta_\odot^\textrm{fid})^2 (\alpha_\odot - \alpha_\odot^\textrm{fid})^2 + (\delta_\odot - \delta_\odot^\textrm{fid})^2}{2 \sigma_\odot^2}
\end{equation}
where $\sigma_\odot = 2 \, \mathrm{arcmin}$ is the uncertainty in the solar position. We also force $B_1,B_2,B_3$ and the data in any ${ijEt}$ bin to be strictly positive. We compute our MCMC chains via the \texttt{emcee} Python package~\cite{emcee}, then extract a limit from the marginalized posterior on $S_0$ using the $\mathrm{CL}_s$ test statistic~\cite{Read_2002}, defined by 
\begin{equation}
    \textrm{CL}_s = \frac{1 – p_{S_0}}{p_{S_0=0}}
\end{equation}
where $p_{S_0}$ denotes the cumulative distribution function at a given value of $S_0$. A $90\%$ $\mathrm{CL}$ upper limit on $S_0$ is set by solving $\mathrm{CL}_s = 0.1$.

\subsection{Diagnostic plots}
\begin{figure}[h!]
    \centering
    \includegraphics[width = 0.45\textwidth]{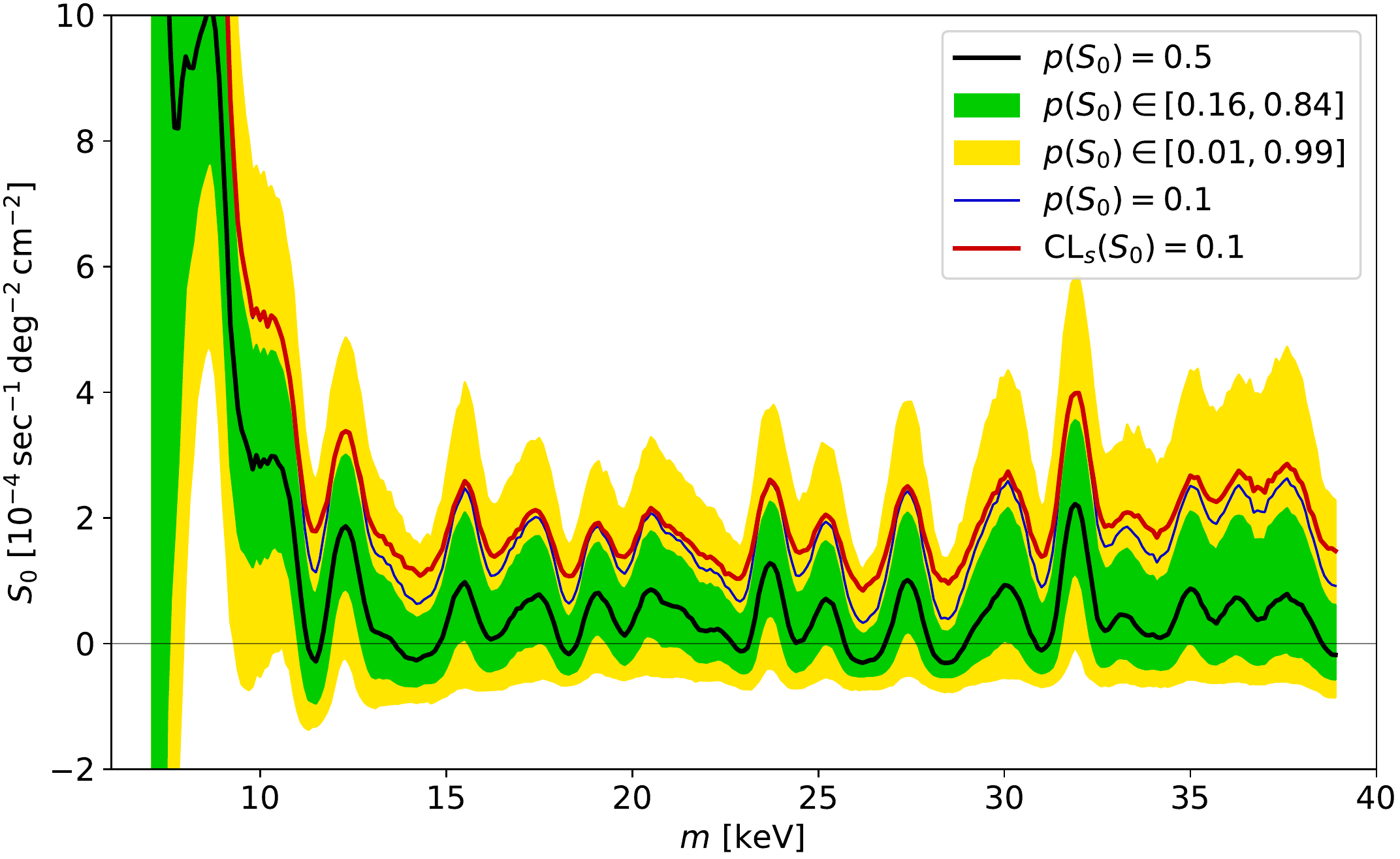}
    \caption{Confidence intervals on $S_0$ using the CL$_s$ method. The bias towards $S_0 > 0$ demonstrates that the model used is in general conservative. \nblink{nb_15_likelihood_data.ipynb}}
    \label{fig:fit_S0}
\end{figure}

Here we include various diagnostic plots for the CL$_s$ fit for a particular value of axion mass, $m = 15$ keV. Fig.~\ref{fig:spectrum} is the ultimate spectral fit of the various background components. It can be seen that the best fit (blue) coincides well with the observed data (black). Fig.~\ref{fig:corner} shows the corner plot for the MCMC fitting procedure. Fig.~\ref{fig:fit_S0} shows the ultimate confidence intervals on $S_0$ as a function of axion mass.

\section{Optimum cuboid method}
\label{app:yellin}

As discussed in Section~\ref{subsec:yellin}, Yellin's original method~\cite{yellin2002} was designed for searches with a signal model with a known one-dimensional distribution, and for data containing very few events. 
This method was generalized~\cite{yellin2007} to searches with a larger number of events by searching for the largest interval (fraction of the data set) which contains no more than certain number of events. In this way, instead of computing the distance in the phase space between any two consecutive events, one can scan the interval size, and find the interval which sees a largest downward fluctuation in the number of events inside it. The method presented in Ref.~\cite{yellin2007} suggests the possibility of coarsely discretizing the data region, which is essential in our generalization of the method to high statistics and to a higher-dimensional data space.

The main challenge for extending these methods to higher dimensional spaces is the rapid growth of computational complexity with the number of dimensions $D$. A naive expectation for the steps required for finding the largest cuboid with a certain number of data points $n$ in a data set with $N$ total points in $D$ dimensions grows as $N^{2D-2} n$, while more advanced algorithms have reduced the complexity for finding the largest empty cuboid to $\mathcal{O}(N^{8/3} (\log N)^3)$ in 3 dimensions~\cite{Ku1997,NANDY199811}. For our data in $D=4$, $N$ ranges from $\mathcal{O}(10^2)$ at high axion masses to $\mathcal{O}(10^4)$ at low axion masses, where the complexity in the naive prescription is beyond our computational power.

\begin{figure*}
\includegraphics[width = 0.85 \textwidth, trim = 0 370 0 0, clip]{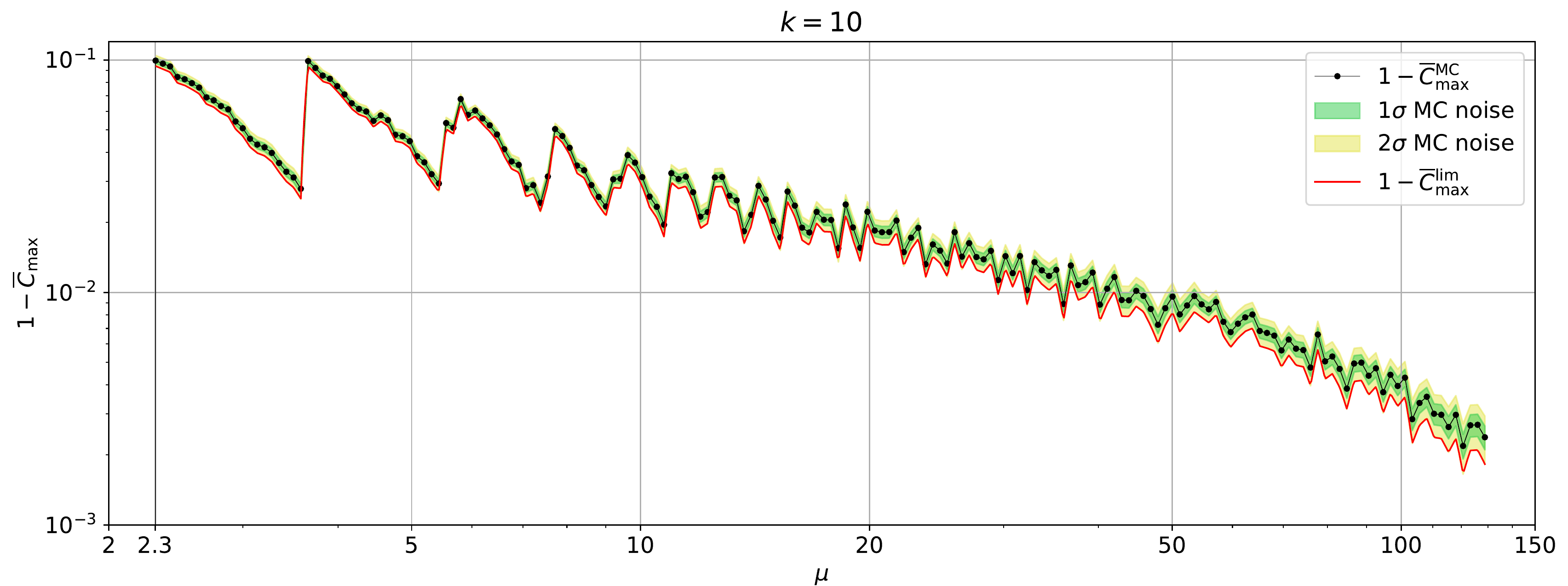}
\caption{The function (one minus) $\overline{C_{\rm max}}$ as a function of the number of signal events as determined by Monte Carlo simulations (black points). Since $\overline{C}_{\rm max}$ is computed from cumulative distribution functions, it has a statistical error which we estimate as $[N_\mathrm{MC}(1-\overline{C}_{\rm max}^\mathrm{MC})]^{1/2}$, shown as $1$- and $2$-sigma bands in green and yellow. This error grows with increasing $\mu$, but we compensate with a larger number $N_\mathrm{MC}$ of samples such that the fractional uncertainty in $1-\overline{C}_{\rm max}$ is $\lesssim 10\%$ for all $\mu$. For limit-setting purposes, we conservatively use the red line $\overline{C}_\mathrm{max}^\mathrm{lim}$, which is its 2-sigma upward fluctuation. \nblink{nb_01_yellin_cmax_bar.ipynb}} \label{fig:cmaxbar}
\end{figure*}

As a result, we generalize the Yellin method to higher dimensions by discretizing the full experimental range into $k^4 = 10^4$ cuboids, constructed such that each cuboid contains exactly $1/k^4$ of the expected signal events. We then generate large number of MC samples for a fixed average of $\mu$ signal events (uniformly distributed on the unit cuboid). We count the number of signal events in each unit cuboid and locate the largest cuboid $V_n^{(i)}$ which contains $n$ number of events for the $i$th MC sample. This list $\lbrace V_n \rbrace^{(i)}$ over the many MC samples is then used to compute the cumulative distribution function of the $V_n$, denoted $C_n(V)$, for each $n$.  In MC sample $i$, the maximum value of $C_n(V)$ achieves for any particular $V_n^{(i)}$ of that sample is denoted $C_\text{max}^{(i)} = \max_n[C_n(V_n^{(i)})]$. A typical $C_n(V_n^{(i)})$ is around 0.5 by construction, as this corresponds to the median. A large $C_n(V_n^{(i)})$ indicates an upward fluctuation of a cuboid volume containing $n$ points in sample $i$. Hence $C_\text{max}^i$ is the least probable downward fluctuation in signal events in the sample. The list $C_\text{max}^{(i)}$ can be used to compute $\overline{C}_{\rm max}$, the value for which the cumulative distribution function of $C_\text{max}^{(i)}$ over all $N_{\rm sim}$ samples reaches $90\%$. The procedure is carried out for a list of $\mu$ values corresponding to signal strength with which we obtain a function $\overline{C}_{\rm max}(\mu)$, plotted in Fig.~\ref{fig:cmaxbar}, which we can compare with data. (A glossary of these terms is provided at the end of this subsection for reference.)

We project each photon observed by NuSTAR, with coordinates ($\alpha$, $\delta$), arrival time, and energy, onto a unit 4D cube. The projection is such that a \emph{pure signal} distribution (i.e.~Eq.~\ref{eq:S0}) would be uniformly distributed on this unit 4D cube; we follow the projection implementation of Ref.~\cite[App.~C]{yellin2007}. We then subdivide this unit cube into $k^4 = 10^4$ bins as we did for the MC samples. For each axion mass $m$ and each of the 317 solar positions samples, this procedure results in a list of events indexed by four integers indicating the generalized bin to which each event belongs. This list can then be used to compute a list of $V^{\rm data}_n$, as well as $C_n ( V^{\rm data}_n)$ of this list of $V_n$ with the cumulative distribution functions obtained from the MC simulations for each $\mu$. We can then find $C^{\rm data}_{\rm max}(\mu) $ as the largest of  $C_n ( V^{\rm data}_n) $ of each $\mu$ value and compare it to $\overline{C}_{\rm max}^\mathrm{lim}(\mu)$ of Fig.~\ref{fig:cmaxbar} for each $\mu$ value.

We have generated MC samples only for $\mu \leq 128$. If $C^{\rm data}_{\rm max}(\mu) < \overline{C}_{\rm max}^\mathrm{lim}(\mu)$ even for $\mu = 128$, we employ a down-sampling strategy. In that case, we randomly select $1/2$ of the events projected on the 4D unit cube, and attempt to set a limit as outlined above on half of this data set. If $\mu = 128$ is still not excluded, we down-sample by another factor of $1/2$, and so forth, in an iterative way. When this protocol converges after $d$ down-sampling steps to a limit $\mu^\mathrm{lim}_{(d)} < 128$, we take the limit on $\mu$ for the raw data to be $\mu^\mathrm{lim} = 2^d \mu^\mathrm{lim}_{(d)}$. Finally, we take the least stringent limit of the 317 solar positions at each axion mass.
The same process is carried out for all axion masses to find the $90\%\mathrm{CL}$ limit $\mu^\mathrm{lim}$ as a function of axion mass. This limit is plotted in the bottom panel of Fig.~\ref{fig:fluxlim} as a red line, with a corresponding flux per unit solid angle at the solar center depicted in the top panel.

A list of the definition of the quantities we used to set limits are:
\begin{itemize}
    \item $k$: Number of generalized bins per MC sample, in each dimension.
    \item $N^{(i)}$: Total number of simulated events in an MC sample, where $i$ indexes the different samples.
    \item $N_\text{MC}$: Number of MC samples generated.
    \item $V_n^{(i)}$: The largest volume that contains no more than $n$ points in the $i$-th MC sample.
    \item $\{V_n\}^i$: The array of $V_n^{(i)}$ for $n = 0,\dots,N^{(i)}$ in sample $i$. Note that $V_{N^{(i)}}^{(i)} = 1$ because the largest volume that contains no more than $N^{(i)}$ points is the entire box.
    \item $\mu = \braket{N^{(i)}}$: Expected number of data points in the MC samples. The following quantities will depend on $\mu$ implicitly.
    \item $P_n(V) = \text{PDF}_{V_n}(V)$: The probability distribution function, \textit{for a particular} $n$, over the $V_n^(i)$ of the MC samples. The location of the peak of this PDF increases monotonically with $n$, since the corresponding average largest volume containing no more than $n$ points necessarily gets larger.
    \item $C_n(V) = \text{CDF}_{V_n}(V)$: The cumulative distribution function of $V_n$, or simply $\int_0^V \dd\tilde{V}\, P_n(\tilde{V}) $. The point at which this curve rapidly saturates towards 1 increases monotonically with $n$, see above.
    \item $C_\text{max}^{(i)}$:  In MC sample $i$, the maximum value that any $C_n(V)$ achieves for the particular $\{V_n\}^{(i)}$ of that sample: $C_\text{max}^{(i)} = \max_n[C_n(V_n^{(i)})]$. Most $C_n(V_n^{(i)})$ would be expected to be around 0.5, as this corresponds to an average MC sample, hence a large $C_n(V_n^{(i)})$ indicates an upwards fluctuation in the large volume containing $n$ points in sample $i$, $C_\text{max}^{(i)}$ is the least likely such fluctuation across all $n$ in the sample. 
    \item $\Upsilon(C) = \text{PDF}_{C_\text{max}} (C)$: The probability distribution function of $C_\text{max}$ over all $N_\text{MC}$ samples.
    \item $\overline{C}_\text{max}$: The value at which the CDF of $\Upsilon(C)$ reaches 0.9, i.e.~$\int_0^{\overline{C}_\text{max}} \dd C \, \Upsilon(C) = 0.9$. 
    \item $C_\text{max}^\text{data}$: This is the maximum of $C_n(V_n)$ across $n$ on the observed data. The $\mu$ value for which $C_\text{max}^\text{data} = \overline{C}_\text{max}$ (note that both sides depend on $\mu$) this corresponds to our a $90\%\mathrm{CL}$ limit $\mu = \mu^\text{lim,Yellin}$.
\end{itemize}

\section{Additional limit plots}
\label{sec:moreplots}

In this section, we include a few additional constraint figures. Figure~\ref{fig:fluxlim} is our ``raw'' exclusion figure upon which all others are based. It shows the limits on $S_0$, the signal flux per unit solid angle at the center of the Sun, as a function of mass, for the three methodologies. The bottom panel shows the corresponding total number of events $\mu$, in comparison to the total number $N$ of events recorded.

\begin{figure}
\includegraphics[width = 0.45 \textwidth]{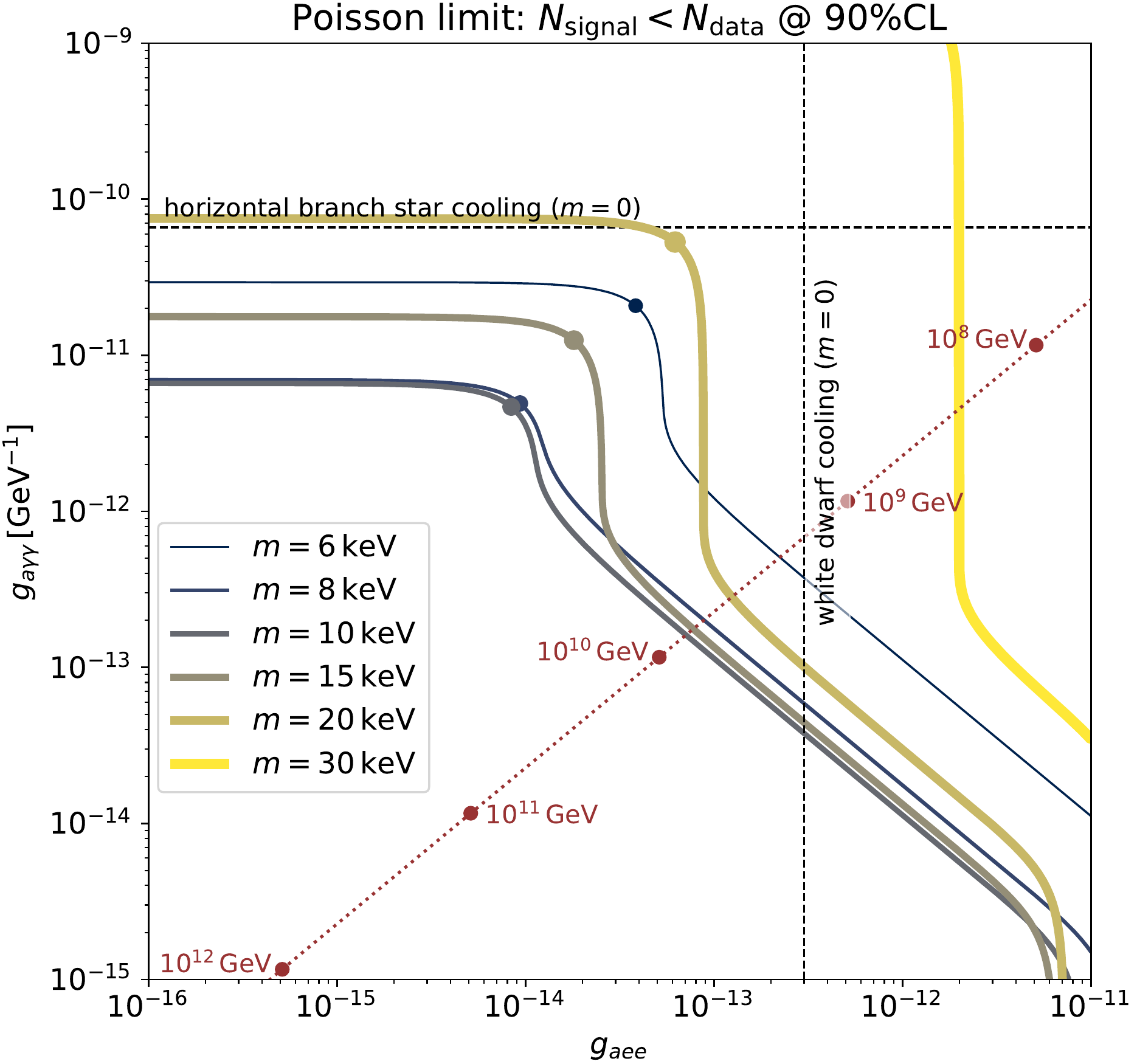}
\includegraphics[width = 0.45 \textwidth]{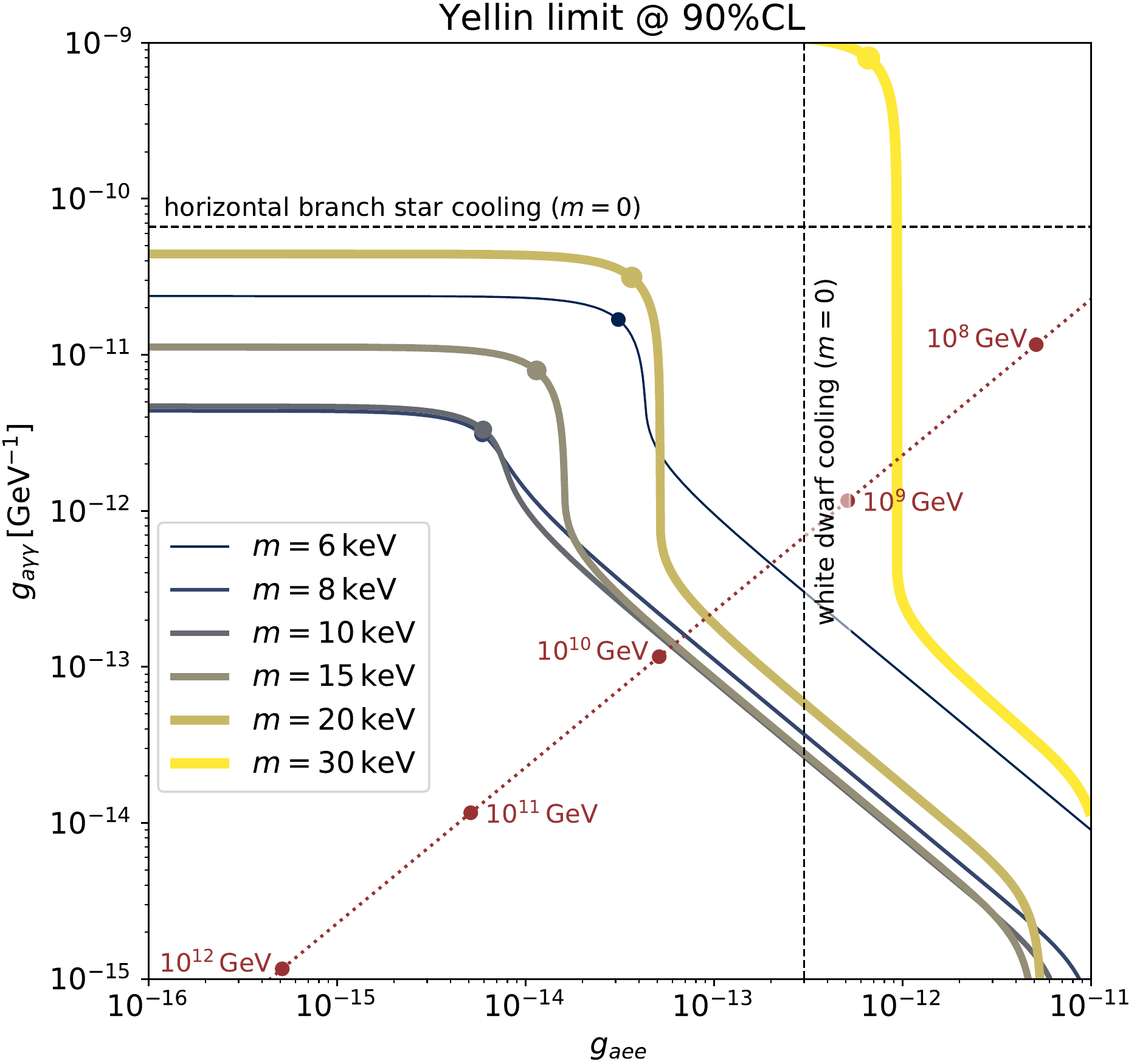}
\caption{The Poisson and Yellin limits in the full three-dimensional parameter space. Each yellow/brown curve corresponds to a particular axion mass, while black dashed lines indicate stellar cooling bounds on the couplings. The red dotted line is the slice corresponding to the ``universal coupling'' displayed in Fig.~\ref{fig:universal}. See text of App.~\ref{sec:moreplots} for details. \nblink{nb_16_lim_data.ipynb}} \label{fig:contourspy}
\end{figure}

\begin{figure}
    \centering
    \includegraphics[width = 0.45 \textwidth]{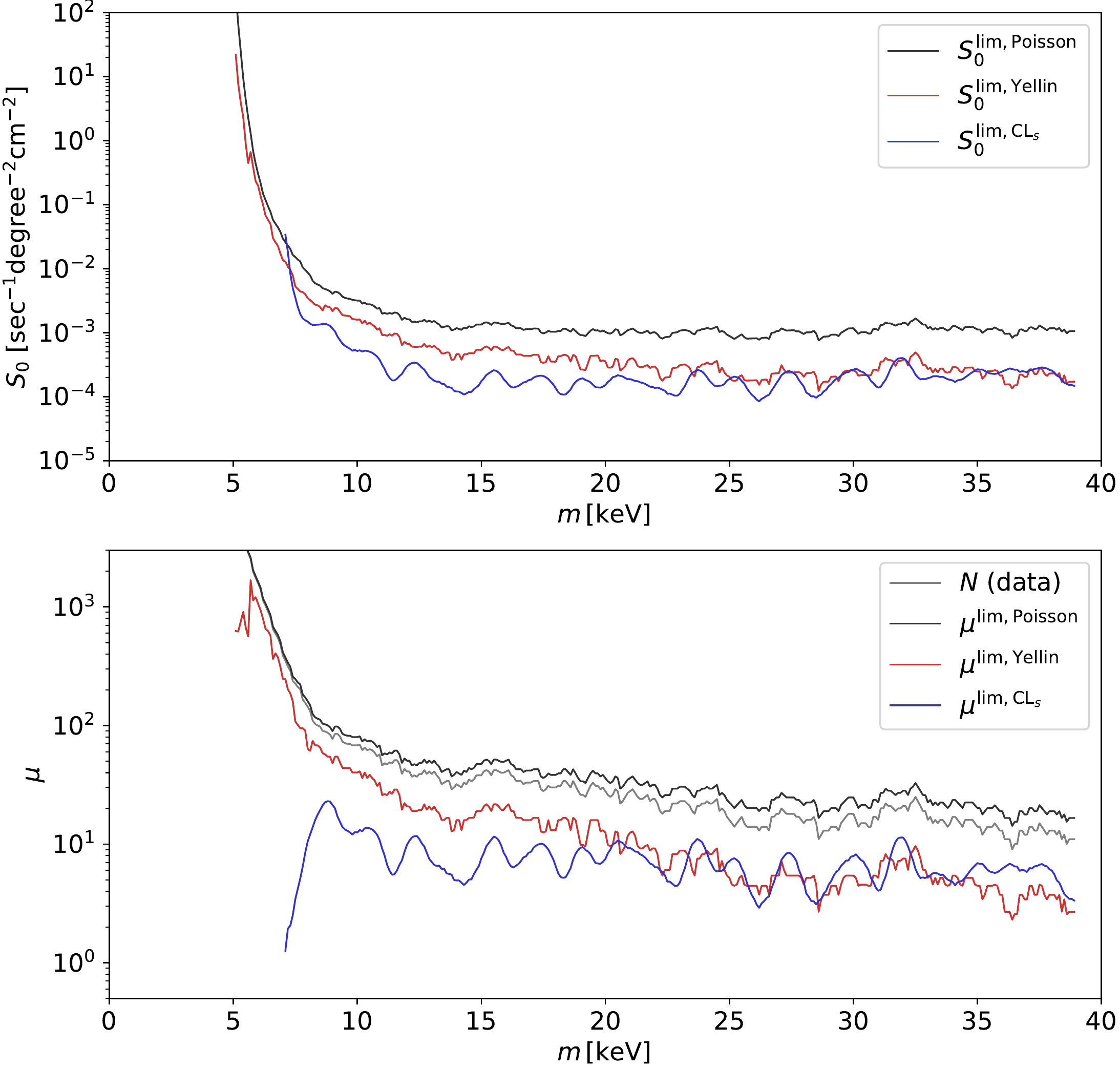}
    \caption{\textit{Top panel:} $90\%\mathrm{CL}$ limits on the signal flux per unit solid angle at $\theta$ = 0, i.e.~$S_0$ from Eq.~\ref{eq:S0}, for the three methodologies (Poisson, Yellin, $\mathrm{CL}_s$; black, red, blue) used in this work. \textit{Bottom panel:} Corresponding number of total expected photon counts $\mu$ for these three limit curves. Also plotted in gray is the total number of observed photons with $E \in [m/2-2\sigma_E, m/2+2\sigma_E]$. \nblink{nb_16_lim_data.ipynb}}
    \label{fig:fluxlim}
\end{figure}

In Fig.~\ref{fig:contourspy}, we show the Poisson and Yellin limits in the full three-dimensional parameter space, the analogs of Fig.~\ref{fig:contours} for the $\mathrm{CL}_s$ limit. The limit curves have various turning points on this three-dimensional parameter space. As an example, consider the $m= 15\,{\rm keV}$ line: going from left to right, the four distinct sections of the curve correspond to when
\begin{itemize}
    \item The axion-photon coupling determines the production and decay rates of the axion (horizontal). 
    \item The axion-electron coupling determines the axion production rate and the axion-photon coupling determines the axion decay rate, and the axion lifetime is shorter than the age of the universe (vertical).
    \item The axion-electron coupling determines the axion production rate while the axion-photon coupling determines the axion decay rate, and the axion lifetime is longer than the age of the universe (oblique).
    \item The axion-electron coupling determines the axion production and decay rates (decay proceeds via an electron loop), and the lifetime of the axion is longer than the age of the universe (vertical).
\end{itemize}
The second and fourth section may not show up for some axion masses.

\end{document}